\documentclass[twocolumn]{aastex701}
\usepackage{apjfonts}
\usepackage{rotating}
\usepackage{amsmath, color,graphicx,tablefootnote,subfigure}
\usepackage{graphics,subfigure,hyperref,float}
\usepackage[rightcaption]{sidecap}

\newcommand\hii{\ion{H}{2}}
\newcommand\hei{\ion{He}{1}}
\newcommand\heii{\ion{He}{2}}
\newcommand\nii{[\ion{N}{2}]}
\newcommand\oi{[\ion{O}{1}]}
\newcommand\oii{[\ion{O}{2}]}
\newcommand\oiii{[\ion{O}{3}]}
\newcommand\neiii{[\ion{Ne}{3}]}
\newcommand\sii{[\ion{S}{2}]}
\newcommand\siii{[\ion{S}{3}]}
\newcommand\cliii{[\ion{Cl}{3}]}
\newcommand\cliv{[\ion{Cl}{4}]}
\newcommand\ariii{[\ion{Ar}{3}]}
\newcommand\ariv{[\ion{Ar}{4}]}
\newcommand\kiv{[\ion{K}{4}]}
\newcommand\feii{[\ion{Fe}{2}]}
\newcommand\feiii{[\ion{Fe}{3}]}
\newcommand\feiv{[\ion{Fe}{4}]}
\newcommand\fev{[\ion{Fe}{5}]}
\newcommand\te{$T_e$}
\newcommand\den{$n_e$}
\newcommand\yp{$Y_{\rm p}$}

\newcommand{\multrow}[1]{\begin{tabular}{@{}c@{}} #1 \end{tabular}}

\shorttitle{The LBT Y$_{\rm p}$ Project VI}
\shortauthors{Rogers et al.}

\begin{document}

\title{The LBT Y$_{\rm p}$ Project VI: Gas-Phase Chemical Abundance Patterns of Metal-Poor Nebulae}

%
%
\author[0000-0002-0361-8223]{Noah S.\ J.\ Rogers}
\affiliation{Center for Interdisciplinary Exploration and Research in Astrophysics (CIERA), Northwestern University, 1800 Sherman Avenue, Evanston, IL, 60201, USA}
\email{noah.rogers@northwestern.edu}
\author[0000-0003-0605-8732]{Evan D.\ Skillman}
\affiliation{Minnesota Institute for Astrophysics, University of Minnesota, 116 Church St. SE, Minneapolis, MN 55455}
\email{skill001@umn.edu}
\author[0000-0003-1435-3053]{Richard W.\ Pogge}
\affiliation{Department of Astronomy, The Ohio State University, 140 W 18th Ave., Columbus, OH, 43210}
\affiliation{Center for Cosmology \& AstroParticle Physics, The Ohio State University, 191 West Woodruff Avenue, Columbus, OH 43210}
\email{pogge.1@osu.edu}
\author[0009-0006-2077-2552]{Erik Aver}
\affiliation{Department of Physics, Gonzaga University, 502 E Boone Ave., Spokane, WA, 99258}
\email{aver@gonzaga.edu}
\author[0000-0003-4912-5157]{Miqaela K.\ Weller}
\affiliation{Department of Astronomy, The Ohio State University, 140 W 18th Ave., Columbus, OH, 43210}
\email{weller.133@buckeyemail.osu.edu}
\author[0000-0002-4153-053X]{Danielle A. Berg}
\affiliation{Department of Astronomy, The University of Texas at Austin, 2515 Speedway, Stop C1400, Austin, TX 78712, USA}
\email{daberg@austin.utexas.edu}
\author[0000-0001-8483-603X]{John J. Salzer}
\affiliation{Department of Astronomy, Indiana University, 727 East Third Street, Bloomington, IN 47405, USA}
\email{josalzer@iu.edu}
\author[0000-0002-2901-5260]{John H. Miller, Jr.}
\affiliation{Minnesota Institute for Astrophysics, University of Minnesota, 116 Church St. SE, Minneapolis, MN 55455}
\email{mill9614@umn.edu}
\author[0009-0009-2024-9317]{Jayde Speigel}
\affiliation{Department of Astronomy, The Ohio State University, 140 W 18th Ave., Columbus, OH, 43210}
\email{spiegel.85@buckeyemail.osu.edu}
\author[0000-0001-6369-1636]{Allison L. Strom}
\affiliation{Center for Interdisciplinary Exploration and Research in Astrophysics (CIERA), Northwestern University, 1800 Sherman Avenue, Evanston, IL, 60201, USA}
\affiliation{Department of Physics and Astronomy, Northwestern University, 2145 Sheridan Road, Evanston, IL, 60208, USA}
\email{allison.strom@northwestern.edu}


\begin{abstract}

Through a combined rest-optical and NIR observing campaign, the Large Binocular Telescope Primordial He Abundance (\yp) Project has delivered a precision measurement of \yp. Given the stringent requirements to measure gas-phase He abundances, the optical spectra contain emission features from a plethora of metals in the Interstellar Medium (ISM). We present gas-phase N, O, Ne, S, Cl, Ar, K and Fe abundance patterns in 62 low-metallicity nebulae. We assess ionization correction factors for each element and select those that reproduce the observed ionic abundance trends. The Ne/O, S/O, and Ar/O abundances do not evolve with metallicity and agree with the solar ratios. Given the Type Ia enrichment of S and Ar, the solar S/O and Ar/O suggest extended star formation histories in local galaxies relative to high-\emph{z} galaxies at fixed O/H. We report the first ISM K abundances in low-metallicity extragalactic nebulae; both Cl/O and K/O are constant with O/H, indicating the production of these odd-Z elements follows that of $\alpha$ elements. Consistent with prior analyses, gas-phase Fe/O is observed to decrease with O/H due to dust depletion; correcting for this produces an average [Fe/O]$=$$-$0.80, implying Fe-deficient stars capable of producing hard ionizing spectra. Finally, the metallicity-insensitive log(N/O) plateau from 50 nebulae is $-$1.43$\pm$0.10, consistent with previous works. Five galaxies have N/O significantly enhanced relative to the primary plateau, and we discuss their properties, mechanisms that could bias measurements of N/O, and nucleosynthetic progenitors that can produce enhanced N at low metallicity.

\end{abstract}


\section{Introduction}\label{s:intro}

The chemical composition of the Interstellar Medium (ISM) reflects the accumulated processes that affect galaxy growth and evolution. Elements heavier than H and He are synthesized in stars that, in turn, return these metals back to the ISM. The total metallicity (e.g., O/H) is sensitive to the net star formation, as well as processes such as dilution from pristine gas inflow or ejection via galactic winds. Alternatively, relative metal abundances (e.g., N/O, Fe/O) are sensitive to the different nucleosynthetic production sites and their timescales of enrichment. For example, elements synthesized in massive stars are returned to the ISM through Type II supernovae (SNe) in $\sim$10 Myr, while metals produced from Type Ia SNe undergo significantly longer time delays \citep[greater than 100 Myr,][]{koba2009}. Therefore, the star-formation history (SFH) can shape the observed chemical abundance patterns in the ISM \citep{koba2020b}.

In contrast to the metals, He has a stellar enrichment component and a primordial origin from Big Bang nucleosynthesis (BBN), where the latter accounts for the majority of the He present in the ISM \citep{peebles1966,koba2020b,weller2025}. The primordial He abundance, \yp, is sensitive to early-Universe physics that alters the production of He during BBN, such as the effective number of neutrino flavors and the baryon-to-photon ratio \citep[e.g.,][]{cyburt2016,fields2020}. Measuring \yp\ to the precision necessary to evaluate the presence of non-Standard Model physics is challenging owing to the systematics associated with measuring He/H from the gas \citep{oliv2001,olive2004,aver2011}. Observationally, historically, \yp\ is constrained via the intercept of the He/H-O/H relation measured in low-metallicity nebulae \citep{peimbert1974,peimbert1976}. This approach would, ideally, leverage a statistical sample of He abundances in extremely metal-poor nebulae, but prior inhomogeneous samples (i.e., using different telescopes, detectors, and techniques to measure O/H or He/H) have resulted in relatively few robust He/H measurements in galaxies with O/H $<$ 0.1$\times$(O/H)$_\odot$ \citep[e.g.,][]{aver2015}.

To address this challenge, the Large Binocular Telescope (LBT) \yp\ Project has conducted a spectroscopic observing campaign of 62 local low-metallicity nebulae. As described in \citet[][hereafter Paper I]{skillman2026}, the sample was selected for observation based on prior measurements of the gas-phase O abundance (also referred to as metallicity in this work) and high surface brightness and EW of the H$\beta$ emission line. Each target was observed with the Multi-Object Double Spectrographs \citep[MODS,][]{pogg2010} to obtain spectroscopic coverage of the rest-optical \ion{He}{1} recombination lines. Using these data, \citet[][hereafter Paper II]{rogers2026yp} measured the H and He line fluxes, gas conditions such as the electron temperature (\te) and density (\den), and robust O/H abundances. In addition, \citet[][hereafter Paper III]{weller2026} analyzed LBT Utility Camera in the Infrared (LUCI) observations of 48 targets in the LBT \yp\ Project sample. The NIR spectra are necessary to measure the density-sensitive \hei$\lambda$10830 emission line, which allows for the most robust measurements of the He$^+$ abundance \citep[e.g.,][]{izotov2014,aver2015,aver2021}. With the homogeneously observed, reduced, and analyzed data, \citet[][hereafter Paper IV]{aver2026} determined the He abundance in each target, assessed systematics related to the He abundances and methodology for measuring \yp, and presented an empirical measurement of \yp\ with a precision of 0.5\%. \citet[][hereafter Paper V]{yeh2026} discuss the cosmological implications for this measurement, where the new value of \yp\ is in good agreement with Standard BBN.

Measurements of gas-phase He abundance require high-quality optical and NIR spectroscopy sensitive enough to detect faint H and He recombination lines.
As a consequence, the LBT/MODS spectra contain many other collisionally-excited lines (CELs) from a variety of metals in the ISM, providing a means to measure metal abundance patterns in the low-metallicity gas.
The pioneering works of \citet{thua1995, izotov1999, izotov2004,izot2006} determined the relative abundances of C, N, O, Ne, Si, S, Ar, and Fe from increasingly large samples of low metallicity galaxies. A primary conclusion from their work was that the abundances of these elements, relative to O, showed no trend with oxygen abundance. This led to the proposal that these abundance were all primary in nature and a result of the yields of Type II SNe integrated over a nearly universal IMF.

Chemical abundance patterns in the ISM have been the subject of renewed focus as new facilities and larger samples have become available. For instance, the DEep Spectra of Ionised REgions Database \citep[DESIRED,][]{mend2023} compiled emission line fluxes for more than 2000 ionized nebulae and homogeneously-analyzed these spectra to measure chemical abundances of $\alpha$-process elements \citep{este2025}, odd-Z elements \citep[such as Cl,][]{orte2025}, and Fe \citep{mend2024}. The second data release of the Dark Energy Spectroscopic Instrument \citep[DESI,][]{desi2024} survey produced a sample of nearly 50000 star-forming galaxies with direct-method \te, \den, and metal abundances \citep{scholte2026}. Both the DESIRED and DESI samples indicate that most $\alpha$ element abundance ratios (e.g., Ne/O) agree with the solar ratios at low O/H. In contrast, observations of high-\emph{z} galaxies with JWST have revealed unexpected ISM abundance patterns, such as enhanced N/O that is inconsistent with the trends of local low-metallicity galaxies \citep[e.g.,][]{bunk2023,topp2024,topping2025,berg2026b}, and significantly sub-solar S/O and Ar/O \citep[e.g.,][]{roge2024,rogers2026cecilia,bhattacharya2025,stan2025,foley2026}. These trends may indicate more complex production mechanisms for these elements, although it remains unclear whether the same mechanisms can be invoked to match the chemistry in similarly metal-poor, highly star-forming galaxies in the local Universe.

With the MODS spectra acquired from the LBT \yp\ Project, we have an opportunity to measure the chemical abundance trends in prototypical low-metallicity star-forming nebulae to high precision. These provide crucial insight into the enrichment sources in metal-poor environments and context for interpreting the emerging trends of high-\emph{z} galaxies.
In \S\ref{s:data} we review the spectroscopic data of the LBT \yp\ Project, the available physical conditions measured from the MODS data, and the ionic abundances measured as part of this analysis. In \S\ref{s:icfs} we compare numerous ICF predictions to the ionic abundance trends measured in the sample, select the ICFs use for the remainder of the analysis, and determine the total metal abundances. We discuss the chemical abundance patterns of eight elements in the low-metallicity ISM in \S\ref{s:abun}, and we further compare these abundances to the derived He abundance in \S\ref{s:heabun}. In \S\ref{s:highno}, we explore the physical properties of five galaxies in the sample with elevated N/O at low O/H, an anomalous abundance pattern in local nebulae, and examine potential sources of the N/O enhancement. We summarize our conclusions in \S\ref{s:conclusions}. Throughout this manuscript, we assume solar abundance patterns from \citet{aspl2021}: 12+log(O/H)$_\odot$ $=$ 8.69$\pm$0.04, log(N/O)$_\odot$ $=$ $-$0.86$\pm$0.08, log(Ne/O)$_\odot$ $=$ $-$0.63$\pm$0.06, log(S/O)$_\odot$ $=$ $-$1.57$\pm$0.05, log(Cl/O)$_\odot$ $=$ $-$3.38$\pm$0.20, log(Ar)$_\odot$ $=$ $-$2.31$\pm$0.11, log(K/O)$_\odot$ $=$ $-$3.62$\pm$0.05, and log(Fe/O)$_\odot$ $=$ $-$1.23$\pm$0.06. We use [X/H] to represent log(X/H) $-$ log(X/H)$_\odot$.

\section{Data and Ionic Abundances}\label{s:data}

\subsection{Spectroscopic Data and Physical Conditions}\label{s:datarecap}

The target selection and data reduction methods have been described in other LBT \yp\ Project papers, and we refer the reader to Papers I, II, and III for these details. In brief, targets are selected for observation based on their low metallicities from prior direct or indirect metallicity studies. Robust He abundances require the high-S/N detection of many H and He recombination lines, so we generally require that F(H$\beta$) $>$ 10$^{-15}$ erg s$^{-1}$ cm$^{-2}$ and EW(H$\beta$) $>$ 100 \AA. There are additional constraints on a target's redshift, as crucial \ion{He}{1} lines can redshift into known telluric and Galactic absorption features and bias the resulting He abundances.

In total, the LBT \yp\ Project targeted 56 extragalactic nebulae for LBT/MODS observations of the rest-optical spectrum, reanalyzed archival MODS data of seven sources, and incorporated the spectra of Leo P and AGC198691 from the earlier works of \citet{aver2021} and \citet{aver2022}. Rest-NIR spectra were acquired with LBT/LUCI for 48 targets, where we adopt literature NIR fluxes for another 11 targets and the LUCI spectra of Leo P reported in \citet{aver2021}. The MODS data have wavelength coverage from 3300-10000 \AA\ at spectral resolution $R$$\sim$2000, while LUCI spectra cover 0.9-1.35 $\mu$m at $R$$\sim$2100. Exposure times for each target can be found in Paper I. Data are reduced using \textsc{modsCCDRed} \citep{pogg2019} and \textsc{modsIDL} \citep{crox2019} for the MODS spectra, and \textsc{PypeIt} \citep{pypeit:joss_pub} for the LUCI spectra. The total number of galaxies with reduced MODS and LUCI spectroscopic data are 62 and 48, respectively, and we refer the reader to Papers II and III for specifics concerning the emission line and continuum fitting in these data.

To measure the He/H abundance, we use the observed fluxes and equivalent widths (EWs) of up to 15 H and 8 He lines in the MODS and LUCI spectra (see Paper IV). As $E(B-V)$ is determined as part of the He abundance model, the spectra are not corrected for reddening at this point. However, \te\oiii\ is used as a weak prior in this model, and O/H is required to interpret any relation between $Y$ and metallicity.
To obtain \te\oiii, the four strongest Balmer lines in the MODS data are used to simultaneously determine $E(B-V)$ and $a_H$, the stellar Balmer absorption EW. We reddening correct the spectra using the \citet{card1989} attenuation law, and account for Balmer absorption in the four strongest Balmer lines. From the reddening-corrected line intensities, we use the auroral-to-nebular line ratios of \nii, \oii, \oiii, \sii, \siii, and \ariii\ to measure a variety of electron temperatures. \te\oiii, which is measured in every LBT \yp\ Project target, is then passed into the He abundance code. We also measure emission line ratios from \oii, \sii, \cliii, and \ariv\ that directly constrain the electron density. See Paper II for more details on these calculations and the physical conditions measured in the full sample.

\begin{deluxetable*}{lccc}  
\tablecaption{Selected Atomic Data and Ionization Zones \label{t:atomic}}
\renewcommand{\arraystretch}{1.0}
\tablewidth{\textwidth}
\tabletypesize{\footnotesize}
\tablehead{
   \colhead{} & 
   \colhead{Transition} & 
   \colhead{Collision} &
   \colhead{Ionization} \\ [-0.2cm]
   \colhead{Ion}  & 
   \colhead{Probabilities} &
   \colhead{Strengths} &
   \colhead{Zone}}
\startdata
N$^+$ 	& \citet{froe2004}  &  \citet{taya2011}  &  Low  \\ 
\hline
O$^+$ 	& \citet{froe2004}  &  \citet{kisi2009}  &  Low  \\
\hline
O$^{2+}$ 	& \citet{froe2004}  &   \citet{stor2014}  &  High  \\
\hline
Ne$^{2+}$ 	& \citet{froe2004}  &   \citet{mcLa2011}  &  High  \\
\hline
S$^+$   & \citet{irim2005}  & \citet{taya2010}  &  Low  \\
\hline
S$^{2+}$    & \citet{froe2006}  &   \citet{huds2012}    &  Int  \\
\hline
Cl$^{2+}$    & \citet{rynk2019}  &   \citet{butl1989}   &  Int   \\
\hline
Cl$^{3+}$    & \multrow{\citet{mend1982},\\ \citet{elli1984},\\ \citet{kauf1986}}  &   \citet{gala1995}  &  High   \\
\hline
Ar$^{2+}$   &   \multrow{\citet{mend1983b}, \\ \citet{kauf1986}}    &   \citet{gala1995}  &  Int   \\
\hline
Ar$^{3+}$   &   \citet{mend1982}    &   \citet{rams1997}  &  High  \\
\hline
K$^{3+}$   &   \citet{kauf1986}    &   \citet{gala1995}  &  High  \\
\hline
Fe$^{2+}$   &   \multrow{\citet{deb2009}, \\ \citet{mendc2023}}    &   \citet{zhan1996}  &  Low  \\
\hline
Fe$^{4+}$   &   \citet{naha2000}    &   \citet{ball2007}  &  High  \\
\enddata
\tablecomments{Atomic data to measure chemical abundances in the LBT Y$_p$ Project nebulae. Last column includes the assumed ionization zone for the ion.}
\end{deluxetable*}

\subsection{Ionic Abundances}\label{s:ionicabun}

With \te, \den, and the intensity of many metal CELs, it is possible to directly calculate ionic abundances in the ISM. We assume that the ISM is characterized by a three-zone ionization model \citep{garn1992,berg2015}, where ions with similar ionization potential (IP) reside in a single zone with fixed \te\ and \den. Table \ref{t:atomic} summarizes the atomic data and ionization zone for each ion.
We do not further subdivide the ISM into a fourth very-high ionization zone \citep{berg2021}, which contains ions with IP $>$ 54 eV; while a large fraction of the LBT \yp\ Project nebulae have significant nebular \heii\ emission (Paper II), we do not have a reliable temperature diagnostic for this gas from the MODS optical spectra.

The characteristic temperature adopted for each ionization zone depends on the available ions from which \te\ is directly measured from an auroral-to-nebular line ratio (e.g., \oiii$\lambda$4363/\oiii$\lambda$5007). For the high-ionization zone, we use the direct \te\oiii\ in all nebulae. We use \te\siii\ for the intermediate-ionization zone, which is directly measured in 60 nebulae. For the remaining two, we use the new empirical \te\ scaling relation derived from the LBT \yp\ Project nebulae to infer \te\siii\ from the measured \te\oiii\ (Eq 3 in Paper II). This relation has a small intrinsic scatter of 340 K, which is considered when reporting the uncertainty on the inferred \te\siii\ and the resulting ionic abundances in the intermediate-ionization zone. Finally, we prioritize direct \te\nii\ in the low-ionization zone when available, although this is only measured in 20 targets. While \te\oii\ and \te\sii\ are measured in more nebulae, we observe significant scatter in these \te\ with respect to \te\oiii. Due to the many sources of systematic uncertainty associated with \te\oii\ and \te\sii\ \citep[][Paper II]{mend2023}, we follow the strategies of similar chemical abundance surveys and de-prioritize these direct \te\ \citep[see also][]{kenn2003b,berg2020}. Instead, we infer the low-ionization zone temperature from the direct \te\oiii\ and the \citet{garn1992} \te\ scaling relation, and we include an additional 610 K uncertainty term based on the dispersion observed in empirical direct \te\oiii\ and \te\nii\ \citep{roge2021}.

We derive a characteristic electron density for \te\ and ionic abundance measurements using an adapted version of the prioritizations recommended in \citet{mend2023}. When \den\sii\ is less than 100 cm$^{-3}$, the gas is assumed to be in the low-density limit where ionic abundance measurements from the optical CELs are insensitive to \den. In this instance, the adopted density is \den\ $=$ 100$\pm$50 cm$^{-3}$. If \den\sii\ $>$ 100 cm$^{-3}$, then the characteristic density is an average of \den\sii\ and \den\oii\ when both are available. Collisional de-excitation is an important consideration for nebulae with high electron densities and for CELs with low critical densities.
We explored using higher densities from other ions for ionic abundance calculations in the low- and high-ionization zones, respectively, but found that the \te\ and abundances were statistically unaffected (Paper II). This is because all LBT \yp\ Project nebulae have \den\oii\ $<$ 10$^{3}$ and \den\ariv\ $<$ 10$^4$ cm$^{-3}$, such that all relevant CELs are robust to collisional de-excitation.

We use the characteristic electron density for all ionic abundances, with two notable exceptions. For Cl$^{2+}$ and Ar$^{3+}$, we use the direct \den\cliii\ and \den\ariv\ when available, since these CELs have relatively low critical densities which can systematically bias the ionic abundances when using the average of the low-ionization densities from \sii\ and \oii. If \den\ariv\ is directly measured, we use the isolated \ariv$\lambda$4740 emission line to derive Ar$^{3+}$/H$^+$. Otherwise, we use the characteristic density and both \ariv$\lambda$4711 and \ariv$\lambda$4740, where the final adopted Ar$^{3+}$/H$^+$ is taken as an average of the ionic abundance inferred from the two emission lines. Similarly, if \den\cliii\ is constrained then we measure Cl$^{2+}$/H$^+$ from \cliii$\lambda$5517, otherwise we use the characteristic density and average the Cl$^{2+}$/H$^+$ abundances measured from \cliii$\lambda$5517 and \cliii$\lambda$5537 separately.

The low-, intermediate-, and high-ionization zone \te, the characteristic \den, \den\cliii, and \den\ariv\ are taken directly from Paper II. 
To measure ionic abundances relative to H$^+$, we apply the \textsc{PyNeb} \citep{luri2012,luri2015L,morisset2020} \textsc{getIonAbundance} function using the emission line intensity ratio relative to H$\beta$, the measured or inferred ionization zone \te, the characteristic density, and the atomic data given in Table \ref{t:atomic}. We calculate the uncertainty in the measured ionic abundance using the line intensity ratio uncertainty and the error on \te. It is advantageous to directly calculate the relative abundance of specific metal ions in the ISM if, for instance, the two ions have the same IP or if the emissivity of the emission lines have a similar \te\ dependence (e.g., N$^+$/O$^+$). To perform these calculations, we use the metal CEL line intensities along with the \textsc{getEmissivity} function and appropriate ionization zone \te. Finally, for S$^{2+}$ we use the ratio of the \siii\ nebular lines to assess telluric contamination and select the appropriate emission line for S$^{2+}$ ionic abundances.

The ionic abundances calculated for the full sample are provided in Table \ref{t:abun}. Using this approach, we determine the ionic abundance of up to 13 different ions in the ISM (see Table \ref{t:atomic}).

\section{Total Abundances and Ionization Correction Factors}\label{s:icfs}

All relevant ionization states of O emit strong CELs in the rest-frame optical. As discussed in Paper II, we assume that O/H $\approx$ O$^{+}$/H$^+$ + O$^{2+}$/H$^+$. Neutral O$^0$ is assumed to be outside the ionized region, and O$^{3+}$ constitutes a few percent of the total O abundance in similarly high-ionization star-forming galaxies \citep{berg2021,rickardsvaught2025}. To convert the ionic abundances to total metal abundances for the rest of the elements, we require Ionization Correction Factors (ICFs). 
Given the difficulty associated with detecting all ionized species of a metal in the ISM, ICFs are exceptionally difficult to calibrate on empirical direct abundance data \citep[e.g.,][]{este2015,dors2013}. Instead, ICFs are constructed by arguments of similar IPs of the relevant metal ions \citep{peim1969,crox2016} or are calibrated on photoionization model ionic abundance trends \citep{thua1995,izot2006,delg2014,amay2021}. 

Here, we discuss the ICFs selected for the elemental abundances measured in the LBT \yp\ Project nebulae. For an element X, X/H may be measured from the sum of ionic abundances (or an individual ionic abundance),
\begin{equation}\label{e:icf}
    \text{log(}\frac{\text{X}}{\text{H}}\text{)}=\text{log(}\frac{\sum\text{X}^{i+}}{\text{H}^+}\text{)} + \text{log(ICF(X))},
\end{equation}
where ICF(X) is generally dependent on the ionization state of the gas (e.g., O$^{2+}$/O). Rearranging Equation \ref{e:icf} and adopting a fiducial metal abundance, log(X/H)$_{fid}$, the predicted metal ionic abundance can be plotted as a function of the ICF:
\begin{equation}\label{e:revicf}
    \text{log(}\frac{\sum\text{X}^{i+}}{\text{H}^+}\text{)} = 
    \text{log(}\frac{\text{X}}{\text{H}}\text{)}_{fid} - \text{log(ICF(X))}.
\end{equation}
When ICFs are calibrated to correct relative metal ionic abundances, Equation \ref{e:revicf} can be updated to consider a fiducial metal abundance ratio such as log(X/O)$_{fid}$. In the following subsections and figures, we use the predicted ionic abundance trends to select ICFs that reflect the empirical measurements and will not introduce systematic trends in total abundance measurements. For more straightforward comparison, we also prioritize ICFs that have been selected in recent studies focused on chemical abundance patterns in low- and high-metallicity nebulae \citep[e.g.,][]{arel2024,este2025,scholte2026}.

\subsection{Nitrogen}

\begin{figure}[t]
   \centering
   \includegraphics[width=0.47\textwidth]{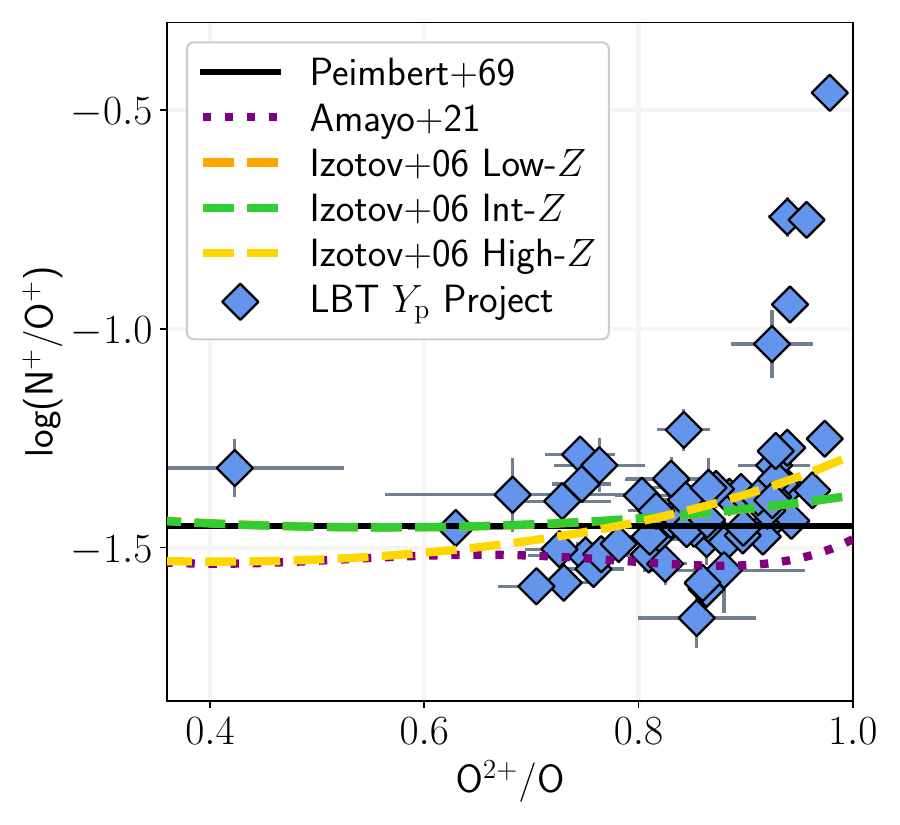}
   \caption{N$^+$/O$^+$ abundances plotted against the O$^{2+}$ ionic fractions in the LBT \yp\ Project nebulae (blue diamonds). Lines represent the ionic abundance predictions vs.\ ionic fraction for different ICFs, offset by a fiducial log(N/O) $=$ $-$1.45. Both the direct abundances and the ICF predictions show minimal evolution with ionization, as may be expected for two ions in the same ionization zone.}
   \label{fig:n_icf}
\end{figure}

The only ion of N with emission lines in the MODS spectra is N$^+$, a low-ionization ion that likely comprises a small fraction of the total N present in the highly-ionized ISM of the typical LBT \yp\ Project target. The IP of N$^+$ and O$^+$ are nearly coincident, which has led many to approximate N/O $\approx$ N$^+$/O$^+$ \citep{peim1969}, or ICF(N) $=$ O/O$^+$. The empirical N$^+$/O$^+$ abundances are plotted as a function of O$^{2+}$/O in Figure \ref{fig:n_icf}, where the ICFs are scaled to a fiducial log(N/O)$_{fid}$ $=$ $-$1.45. This value is selected based on the plateau in N/O abundances observed in other metal-poor nebulae \citep{vanZ2006} and will be discussed further in \S\ref{s:noabun}. The \citet{peim1969} ICF does not vary with ionization, and the LBT \yp\ Project nebulae also show minimal evolution with O$^{2+}$/O. The Pearson correlation coefficient between log(N$^+$/O$^+$) and O$^{2+}$/O is $r$ $=$ 0.16 with $p$ $=$ 0.25 when removing the five targets with log(N$^{+}$/O$^+$) $>$ $-$1.1, or $r$ $=$ 0.35 with $p$ $=$ 0.006 when all targets are considered.
These results indicate that there is no strong trend in the relative N$^+$/O$^+$ abundance with ionization in the LBT \yp\ Project nebulae and that the \citet{peim1969} ICF(N) provides a reasonable fit to the ionic abundances. 

Other studies have attempted to include a correction for the ionization structure of the ISM in ICF(N). Using a suite of photoionization models constructed with low gas-phase metallicity (5\% to 20\% solar metallicity), \citet{nava2006} found that ICF(N) is closer to 1.08$\times$O/O$^+$, consistent with IP-based ICF to within 10\%. The ICFs of \citet{izot2006}, which consider the O/H abundance of the gas to select models ionized by low-, intermediate- and high-metallicity stars, are plotted in Figure \ref{fig:n_icf}, where the low- and intermediate-$Z$ models predict a small increase in log(N$^+$/O$^+$) at O$^{2+}$/O$^+$ $>$ 0.7. \citet{amay2021} developed new ICFs for a number of elements using a grid of photoionization models selected for their ability to reproduce emission line ratios observed in metal-poor star-forming galaxies and metal-rich extragalactic \hii\ regions. The resulting fifth-order polynomial ICF plotted as a dotted purple line in Figure \ref{fig:n_icf} is parameterized as a function of O$^{2+}$/O. This ICF(N) has a maximum value of 1.23$^{+0.66}_{-0.24}\times$O/O$^+$ for nebulae with O$^{2+}$/O $>$ 0.5; while consistent with the commonly-adopted ICF(N) $=$ O/O$^+$, the large uncertainty in the \citet{amay2021} ICF reveals that it may be inappropriate for some nebulae.

One source of systematic uncertainty with the assumption that N/O $\approx$ N$^+$/O$^+$ is the sensitivity of the O$^+$ ionic abundance to \den\ variations. As discussed in Paper II, the large \te\oii\ and \te\sii\ measured in the LBT \yp\ Project sample could be the result of fainter nebular line flux due to collisional de-excitation of the strong \oii\ and \sii\ lines. We use an average of \den\sii\ and \den\oii\ when \den\sii\ is $>$ 100 cm$^{-3}$, and we find that all O$^+$ abundances are consistent when calculated using the direct \den\oii. The \oii\ auroral-to-nebular line ratios can be used to derive \den\ assuming a characteristic low-ionization zone \te\ \citep[see][]{mend2023}, and we reported a higher average \den\oii\ in the LBT \yp\ Project when using these line ratios (see Paper II). Using these higher densities results in a decrease in N$^+$/O$^+$ by $\sim$9\%, similar to the average reported uncertainty on the ionic N$^+$/O$^+$ abundance ratio. In summary, since the change to N$^+$/O$^+$ is generally at the level of the reported uncertainty, we do not correct for potential density biases and adopt the \citet{peim1969} ICF(N) $=$ O/O$^+$.

\subsection{Alpha Elements}

\begin{figure*}[t]
   \centering
   \includegraphics[width=0.98\textwidth]{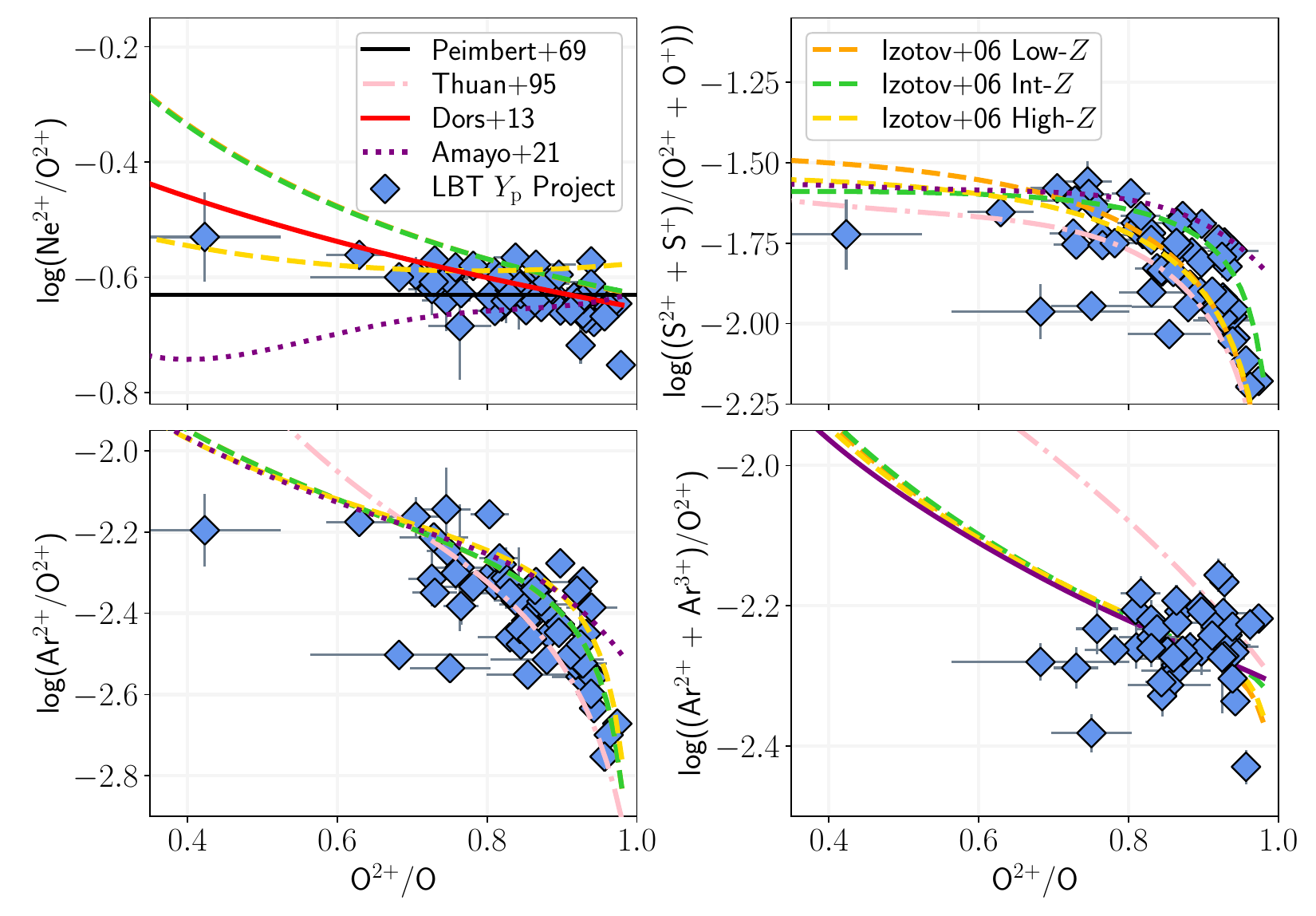}
   \caption{Relative ionic abundances plotted against the O$^{2+}$ ionic fractions (blue diamonds) and ICF predictions (lines) for the $\alpha$ elements Ne, S, and Ar. Each ICF prediction is plotted using the solar abundance ratios as the fiducial abundance (see Equation \ref{e:revicf}). Top Left: Ne$^{2+}$/O$^{2+}$, the \citet{dors2013} ICF best reproduces the observed anti-correlation in the empirical data. Top Right: (S$^{2+}$ + S$^+$)/(O$^{2+}$ + O$^+$), an interpolation of the low- and intermediate-$Z$ ICFs of \citet{izot2006} captures the spread in the observed ionic abundances. Bottom Left: Ar$^{2+}$/O$^{2+}$, the highest ionization nebulae are matched with the \citet{izot2006} ICFs but large scatter remains if Ar$^{3+}$ is not measured. Bottom Right: (Ar$^{2+}$ + Ar$^{3+}$)/(O$^{2+}$), most ICFs predict negligible contributions from Ar$^+$ in the high-ionization ISM. To remain consistent with S and Ar$^{2+}$, the \citet{izot2006} ICFs are adopted.}
   \label{fig:alpha_icf}
\end{figure*}

Emission lines from the $\alpha$ elements Ne, S, and Ar are detected in most of the LBT \yp\ Project nebulae. We plot different ionic abundance ratios and ICF predictions against the O$^{2+}$/O ionic fraction in Figure \ref{fig:alpha_icf}. Although this is not an exhaustive list of all ICFs, those plotted in Figure \ref{fig:alpha_icf} have been extensively utilized by spectroscopic surveys of \hii\ regions and star-forming galaxies \citep[e.g.,][]{berg2020,roge2021,arel2024,este2025,scholte2026}. In all cases, the fiducial abundance ratio used to scale the ICFs is the solar abundance pattern.

Similar to N, only one ionization state of Ne, Ne$^{2+}$, is detected in the optical. While Ne$^{2+}$ is the primary ionization of Ne in high-ionization nebulae, we require an ICF to account for the potentially significant missing Ne$^+$ in nebulae of moderate metallicity/ionization. \citet{peim1969} proposed another ICF based on the IP of O$^{2+}$ and Ne$^{2+}$, where Ne/O $\approx$ Ne$^{2+}$/O$^{2+}$ or ICF(Ne) $=$ O/O$^{2+}$. However, the IP of O$^{2+}$ and Ne$^{2+}$ are not coincident and could result in different ionic fractions in the ISM. In their study of the $\alpha$ element abundance patterns in the COS Legacy Archival Spectroscopy Survey \citep[CLASSY,][]{berg2022,jame2022} galaxies, \citet{arel2024} compared the predicted Ne$^{2+}$/O$^{2+}$ trends of numerous Ne ICFs \citep[e.g.,][]{izot2006,dors2013,amay2021} to directly-measured ionic abundances. They found that the \citet{dors2013} ICF(Ne) produced the lowest dispersion in the resulting Ne/O abundances. \citet{este2025} perform a similar comparison for the DESIRED-Extended \citep[DESIRED-E,][]{mend2024} sample, but they found that the \citet{izot2006} ICFs produce lower dispersion in Ne/O and no secondary trend with ionization, particularly in lower-ionization nebulae.

The Ne$^{2+}$/O$^{2+}$ vs.\ O$^{2+}$/O trends in the LBT \yp\ Project are plotted in the top left panel of Figure \ref{fig:alpha_icf}. Although the majority of the targets are relatively high ionization, there is a clear anti-correlation between Ne$^{2+}$/O$^{2+}$ and O$^{2+}$/O (Pearson correlation coefficient $r$ $=$ $-$0.58, $p$ $=$ 6$\times$10$^{-7}$) that illustrates the increasing importance of O$^+$ and Ne$^+$ in low-ionization nebulae. This is not predicted by the \citet{peim1969} or \citet{amay2021} ICFs, where the latter clearly does not match the measured ionic abundances at O$^{2+}$/O $\lesssim$ 0.7. We choose the \citet{dors2013} ICF(Ne) for this work, which both predicts this anti-correlation and matches the measured ionic abundance trends. This is consistent with the methods used for the high-ionization CLASSY galaxies.

To calculate the abundance of S in the ISM, we use the S$^+$ and S$^{2+}$ abundances measured from the \sii$\lambda$6717,6731 and \siii$\lambda$9069,9532 CELs. While \siii$\lambda$9532 redshifts out of the wavelength coverage of MODS at \emph{z}$\sim$0.049, the \siii$\lambda$9069 line is available for the full sample and is used when required. The ionization energy of O$^{+}$ and S$^{2+}$ are similar, implying a large fraction of missing, highly-ionized S that we cannot measure in the optical/NIR. From the top right panel of Figure \ref{fig:alpha_icf}, the \citet{amay2021} ICF(S) does not match the ionic abundance patterns in the most highly ionized nebulae. This suggests that the fraction of S$^{3+}$ is not properly accounted for and would introduce a spurious trend in S/O with the ionization state of the gas. The \citet{thua1995} ICF(S) captures the shape of the ionic abundances, but is offset to lower ionic abundance ratios. This offset is related to differences in the solar abundance scale adopted in the photoionization models. \citet{thua1995} use the \citet{stas1990} photoionization models that assume log(S/O) $=$ $-$1.40, while \citet{amay2021} use the Bayesian Oxygen and Nitrogen abundance Determination \citep[BOND][]{vale2016} grid which adopts log(S/O) $=$ $-$1.65. The latter is encoded in the photoionization models of \citet{stasinska2015} and is similar to the adopted fiducial S/O used to plot the ICF trends.

The ICF(S) schematic that we adopt for the LBT \yp\ Project is that of \citet{izot2006}. These ICFs match the observed spread in S ionic abundances at fixed O$^{2+}$/O and produce S/O that are uncorrelated with O$^{2+}$/O, a result that has been found in other works \citep{arel2024,este2025}. The functional form of the \citet{izot2006} ICF(S) depends on the gas-phase metallicity and is dependent on O$^+$/O; when the measured 12+log(O/H) is $\leq$ 7.2 or $>$ 8.2, we use the low-$Z$ and high-$Z$ ICF calibrations, respectively. For metallicities at 7.2 $<$ 12+log(O/H) $\leq$ 7.6, we use a linear interpolation of the low- and intermediate-$Z$ ICFs, and a similar interpolation is used for the intermediate- and high-Z ICFs when 7.6 $<$ 12+log(O/H) $\leq$ 8.2 \citep[as recommended by][]{izot2006}.

We detect the rest-NIR \ariii\ emission lines in the majority of targets, as well as the high-ionization \ariv\ CELs in 45 nebulae. We require two ICFs for Ar: one when the only available ionization state is Ar$^{2+}$, and another when both Ar$^{2+}$ and Ar$^{3+}$ are measured. From the bottom panels of Figure \ref{fig:alpha_icf}, we find that the interpolated \citet{izot2006} equations and the \citet{amay2021} relations produce similar Ar/O with low scatter. The shape of these ICFs is consistent when both ions of Ar are available, but the predictions for Ar$^{2+}$/O$^{2+}$ differ.
Additionally, the scatter in Ar/O is significantly larger when Ar$^{3+}$/H$^+$ is missing \citep[see discussion in][]{arel2024}. As the bottom left panel illustrates, the \citet{amay2021} ICF(Ar) does not mirror the Ar$^{2+}$/O$^{2+}$ trends in the highest ionization objects, suggesting that this ICF may be under-predicting the ionic fraction of Ar$^{3+}$ in the high-ionization ISM. 
To remain consistent with the methods of other chemical abundance studies and our ICF(S) selection, we use the \citet{izot2006} ICF(Ar). Ar abundances are calculated using the ICF(Ar) that only corrects for missing Ar$^+$ when possible (i.e., bottom right panel); in cases where the \ariv\ lines are not detected, we must use the ICF(Ar) that corrects for both Ar$^+$ and Ar$^{3+}$.

In deriving uncertainties on total $\alpha$ element abundances, we consider uncertainties in the ionic abundances and the ICF, the latter of which is determined via error propagation of the O ionic fraction uncertainty into the ICFs.
We comment here that the selection of different calibrations of the Ne, S, and Ar ICFs highlights a need for consistent, reliable ICF prescriptions in the low-metallicity, high-ionization ISM. Ideally, these models would incorporate important physical effects that increase the hardness of the ionizing spectrum \citep[e.g., binary evolution,][]{eldr2017} and can match the observed conditions of the extreme ISM observed in these local nebulae and metal-poor high-\emph{z} galaxies. Varying elemental abundance patterns may not change the relative ionic fractions, although this should be verified for ICFs using ionization states of O as a normalization.
Finally, detecting missing metal ionization states is possible from the UV and MIR, enabling empirical constraints on ionic fractions and ICFs \citep[e.g.,][]{dors2013}. This method presents its own challenges owing to differences in UV, optical, and MIR observations, but the capabilities of JWST and future space observatories may enable direct assessments of photoionization model ICFs and, ultimately, more reliable total abundance estimates.

\subsection{Other Trace Elements}\label{s:othericf}

The elements discussed above comprise the typical suite of abundances one could hope to measure in the ISM from an optical/NIR spectrum of a typical star-forming galaxy. The quality and depth of the MODS optical spectra reported here reveal faint CELs from other elements in the ISM that are suitable for chemical abundance measurements. The next most common element detected is Fe, with emission lines from Fe$^+$, Fe$^{2+}$, and Fe$^{4+}$. As noted in previous works, the commonly-detected \feii\ lines have a significant fluorescence component and are unfit for ionic abundance measurements \citep{rodr1999}. Fe$^+$ also resides in the photo-dissociation region and is a very small fraction of the ionized Fe in high-ionization \hii\ regions. We detect \feiii$\lambda$4658 and \fev$\lambda$4227 in 43 and 13 nebulae, respectively.
Unfortunately, the \feiv$\lambda$6740 line is not significantly detected in any of the targets, so we must rely on the available ICFs to estimate the total gaseous Fe abundance.

\begin{figure}[t]
   \centering
   \includegraphics[width=0.47\textwidth]{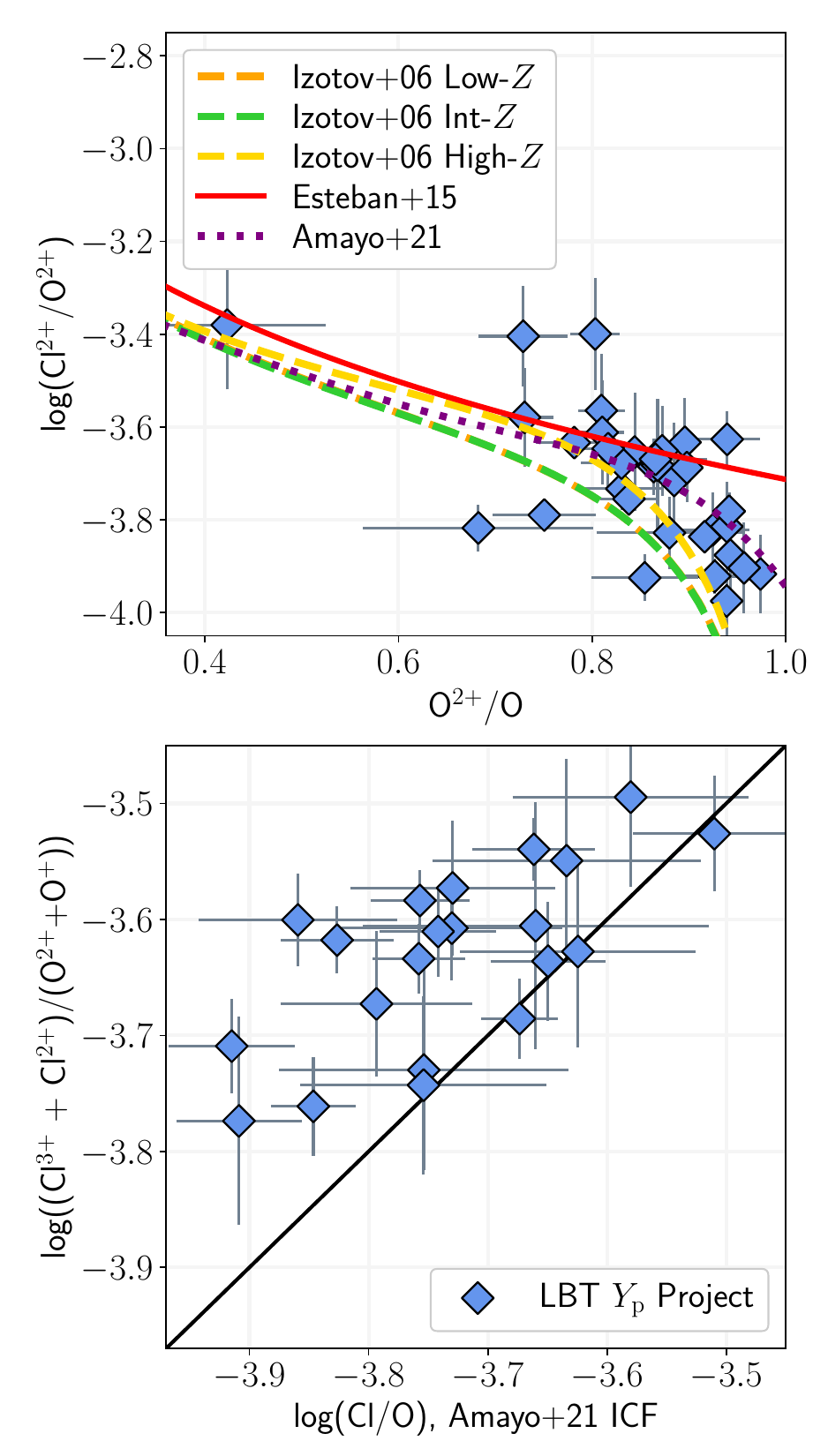}
   \caption{Top: Cl$^{2+}$/O$^{2+}$ vs.\ O$^{2+}$ ionic fraction for the LBT \yp\ Project nebulae (blue diamonds). Different ICF trends are plotted as lines, and the \citet{amay2021} ICF(Cl), offset based on the average Cl$^{2+}$/O$^{2+}$, best reproduces the empirical trends. Bottom: Cl/O derived directly assuming negligible Cl$^+$ (vertical) vs.\ the inferred Cl/O using Cl$^{2+}$/O$^{2+}$ and the \citet{amay2021} ICF(Cl) in the same nebula. One-to-one line is shown in black. There is a systematic offset to higher direct Cl/O, implying that the ICF(Cl) under predicts the contribution of Cl$^{3+}$ in the high-ionization ISM.}
   \label{fig:cl_icf}
\end{figure}

The photoionization model ICFs of \citet{rodr2005} and \citet{izot2006} are often invoked for estimating Fe/H from a measured Fe$^{2+}$/H$^+$ ionic abundance. \citet{mend2024} compared these two ICFs for nebulae with multiple Fe ionic abundances, finding that the \citet{rodr2005} calibration produced better agreement with the summed ionic abundances of Fe. A similar comparison is not possible for the present sample, so we adopt the procedures of \citet{mend2024} and use the \citet{rodr2005} ICF(Fe).
When it is possible to constrain Fe$^{2+}$/H$^+$ and Fe$^{4+}$/H$^+$, we also compare these abundances to those estimated using the ICF(Fe$^{2+}$ + Fe$^{4+}$) of \citet{berg2021}. These ICFs are calibrated for different gas-phase metallicities and are a function of the ionization parameter, which we constrain using O$_{32}$ and the photoionization model relation of \citet{berg2019}. We linearly interpolate between the ICFs based on the direct 12+log(O/H) to determine ICF(Fe$^{2+}$ + Fe$^{4+}$), which accounts for the Fe$^{3+}$/H$^+$ abundance \citep[a significant fraction of the gas-phase Fe at high-ionization, see Figure 7 of][]{berg2021}. We find good consistency between the Fe/O abundances derived using the \citet{berg2021} and \citet{rodr2005} ICFs, so we adopt the latter for a uniform estimation of the Fe abundance.

Emission lines of Cl are detected in 39 LBT \yp\ Project targets: one line of the density-sensitive \cliii$\lambda$5517,37 doublet is detected in 32 nebulae, and 26 targets have at least one of the faint \cliv$\lambda$7540,8046 lines. The detection rate of \cliv\ relative to \cliii\ is due, in part, to the highly-ionized ISM of the sample, but is also a consequence of the lower sensitivity of MODS near the dichroic crossover at $\sim$5700 \AA. Unfortunately, there is no published ICF to solely correct for the missing Cl$^+$ abundance in the ISM, as the majority of the available literature ICFs correct for unobserved Cl$^+$ and Cl$^{3+}$ \citep{izot2006,este2015,amay2021,domi2022}.

The top panel of Figure \ref{fig:cl_icf} plots the Cl$^{2+}$/O$^{2+}$ ionic abundances vs.\ O$^{2+}$ ionic fraction for the LBT \yp\ Project nebulae and predictions for the ICFs. The fiducial log(Cl/O) assumed for this comparison is approximated as the average log(Cl$^{2+}$/O$^{2+}$), which allows us to investigate the shape of ICF(Cl$^{2+}$) relative to the LBT \yp\ Project sample. The empirical ICF(Cl$^{2+}$) of \citet{este2015} reproduces the observed data with O$^{2+}$/O $\lesssim$ 0.9; the decrease in Cl$^{2+}$/O$^{2+}$ in the nebulae with O$^{2+}$/O $>$ 0.9 reflects the increasing fraction of Cl$^{3+}$ in the ISM, which is larger in the LBT \yp\ Project sample compared to the nebulae used to calibrate the \citet{este2015} ICF. The empirical data are best matched by the \citet{amay2021} ICF(Cl$^{2+}$), which does not introduce a trend in the inferred log(Cl/O) with O$^{2+}$/O. 
We adopt the \citet{amay2021} ICF(Cl$^{2+}$) when a direct measurement of Cl$^{3+}$ is unavailable. In doing so, we also use the 16th and 84th percentile ICFs to estimate the uncertainty in ICF(Cl$^{2+}$) when reporting the total Cl/O.
While \citet{orte2025} found that \citet{izot2006} ICF(Cl$^{2+}$) performed better for the DESIRED-E nebulae, the typical ionic fraction of the DESIRED-E nebulae is O$^{2+}$/O $<$ 0.8. The majority of LBT \yp\ Project nebulae have O$^{2+}$/O $>$ 0.8, where the \citet{izot2006} ICF(Cl$^{2+}$) is a strong function of ionization and can introduce spurious trends in Cl/H and Cl/O vs.\ O$^{2+}$/O.

\begin{figure}[t]
   \centering
   \includegraphics[width=0.47\textwidth]{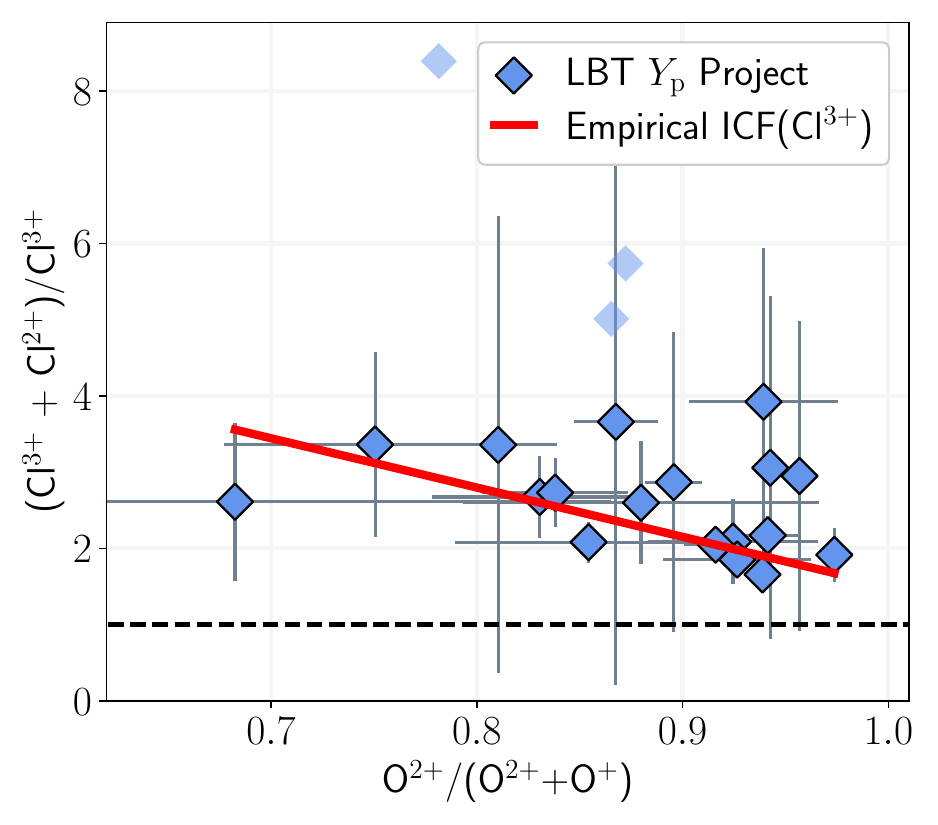}
   \caption{A new empirical calibration of ICF(Cl$^{3+}$) from 18 LBT \yp\ Project nebulae with direct Cl$^{2+}$ and Cl$^{3+}$ abundances. Dashed line indicates no Cl$^{2+}$ in the ISM (i.e., the ICF is negligible). The linear fit in red is provided in Equation \ref{eq:icfcl3}, where targets with poorly constrained Cl/Cl$^{3+}$ fractions (lightly shaded, 3 nebulae) are excluded from the fit. This empirical ICF can be used to infer Cl/H when Cl$^{3+}$ is the only Cl ionic abundance measured in the ISM.}
   \label{fig:icfcl3}
\end{figure}

The Cl$^{3+}$ abundances measured in the high-ionization nebulae present an opportunity to assess the Cl ICFs. If we assume that the abundance of Cl$^+$ is negligible, then the total Cl abundance can be approximated as Cl $\approx$ Cl$^{2+}$ + Cl$^{3+}$ and a direct measurement of Cl/O is readily available. The bottom panel of Figure \ref{fig:cl_icf} plots (Cl$^{2+}$ + Cl$^{3+}$)/O on the vertical axis, while the horizontal axis plots the inferred Cl/O using the Cl$^{2+}$/O$^{2+}$ ionic abundances and the \citet{amay2021} ICF(Cl). There is a systematic trend where the directly-measured Cl/O is larger than the inferred Cl/O by an average of 0.1 dex (maximum difference is 0.25 dex). The approximated direct Cl/O represents the \textit{minimum} Cl/O in the ISM, revealing that the \citet{amay2021} ICFs fail to account for the larger observed ionic fraction of Cl$^{3+}$ in the high-ionization gas. Using this ICF in high-ionization environments will produce scatter to lower Cl abundances, a trend that has been reported in the CLASSY galaxies \citep{arel2024}. Therefore, when Cl$^{3+}$ and Cl$^{2+}$ are measured we assume that the fractional abundance of Cl$^+$ is small, ICF(Cl) $=$ 1, and Cl/O $\approx$ (Cl$^{2+}$ + Cl$^{3+}$)/(O$^{+}$ + O$^{2+}$).

There are eight targets with direct Cl$^{3+}$ abundances that lack Cl$^{2+}$ measurements. No appropriate ICF(Cl$^{3+}$) are available in the literature, but the same Cl $\approx$ Cl$^{2+}$ + Cl$^{3+}$ approximation allows us to measure an empirical ICF from the LBT \yp\ Project nebulae. Figure \ref{fig:icfcl3} plots the Cl/Cl$^{3+}$ vs.\ O$^{2+}$/O ionic fractions for the 21 LBT \yp\ Project targets with direct measurements of both Cl ions. Three nebulae have Cl/Cl$^{3+}$ uncertainties greater than the measured value on account of very low and uncertain Cl$^{3+}$, and these are plotted in light blue. We use Orthogonal Distance Regression (\textsc{ODR}) in \textsc{scipy} \citep{virt2020} to fit Cl/Cl$^{3+}$ as a function of O$^{2+}$/O considering the uncertainties on both variables. The empirical ICF(Cl$^{3+}$) is measured to be
\begin{equation}\label{eq:icfcl3}
    \mbox{ICF(Cl}^{3+}\mbox{)}=(7.97\pm1.60) - (6.47\pm1.73)\times(\mbox{O}^{2+}/\mbox{O}),
\end{equation}
where the uncertainties in the fit parameters are returned from \textsc{ODR}. The new ICF(Cl$^{3+}$) is plotted in red in Figure \ref{fig:icfcl3}; while all nebulae have Cl/Cl$^{3+}$ consistent with the empirical fit, nebulae with relatively uncertain Cl/Cl$^{3+}$ tend to scatter above the best-fit relation. We use Equation \ref{eq:icfcl3} to measure Cl/H and Cl/O in the eight targets with Cl$^{3+}$ ionic abundances only, which all have O$^{2+}$/O $>$ 0.85. Equation \ref{eq:icfcl3} should not be used in nebulae with O$^{2+}$/O $<$ 0.68 (outside the calibration range), and we prefer the combination of Cl$^{2+}$ and Cl$^{3+}$ for an estimate of the total Cl abundance, when available.

\begin{figure}[t]
   \centering
   \includegraphics[width=0.47\textwidth]{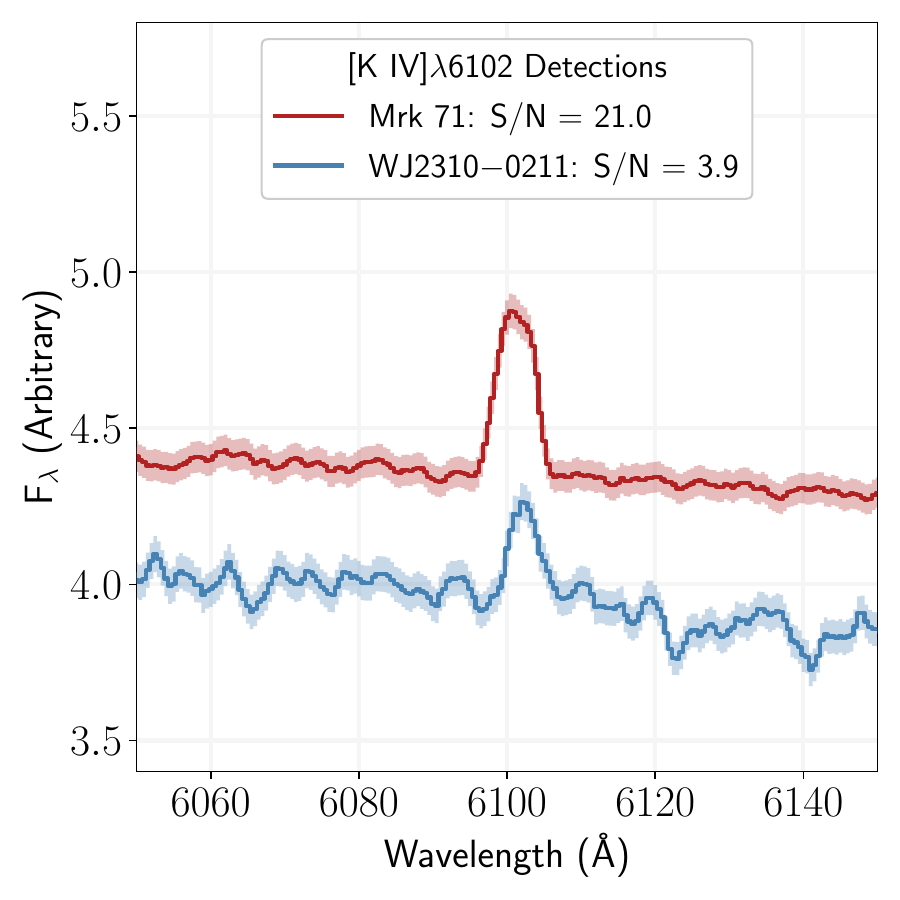}
   \caption{Examples of \kiv$\lambda$6102 detected in low-metallicity extragalactic nebulae. The plotted spectra (errors in shaded area) are from Mrk 71 (red) and WJ2130$-$0211 (blue), which have the highest (21.0) and lowest S/N detection (3.9) of \kiv$\lambda$6102, respectively. There are no other emission features between 6100 and 6105 \AA\ that could feasibly produce the observed emission line.}
   \label{fig:k4}
\end{figure}

A surprising find is the significant detection of the \kiv$\lambda$6102 CEL in ten nebulae. We highlight two examples in Figure \ref{fig:k4}, demonstrating the highest and lowest S/N detections of \kiv$\lambda$6102 in Mrk 71 and WJ2314$-$0211, respectively. This CEL has been detected in numerous planetary nebulae \citep[PNe,][]{alle1978,hyun1994,garc2018} but, to date, has not been reported in the spectra of low-metallicity star-forming galaxies or extragalactic \hii\ regions.
We measure K$^{3+}$/H$^+$ using the high-ionization zone \te\ and \den\sii, but K$^{3+}$ represents only a portion of the K in the ISM.

Unfortunately, there are few available ICFs to measure the total gas-phase K abundance in star-forming galaxies. To infer K/H in the PNe IC 2165, \citet{bohi2013} proposed that K/Ar $\approx$  K$^{3+}$/Ar$^{3+}$ based on the similarity in IP of K$^{3+}$ and Ar$^{3+}$. From grids of PNe photoionization models, \citet{amay2020} found that this simple ICF under predicts the K/Ar abundance by $>$ 0.2 dex, independent of the model ionization state. They calibrated a new ICF(K) as a cubic function of the ionic fraction $\nu$ $=$ He$^{2+}$/(He$^+$ + He$^{2+}$). We detect the emission lines of \heii\ and \ariv\ in all targets with significant \kiv\ emission, permitting us to explore the inferred ICF(K). To do so, we use the \hei\ recombination coefficients of \citet{port2012,port2013} and the intensity of \hei$\lambda$5876, the atomic data of \citet{stor1995} and the intensity of the narrow, nebular \heii$\lambda$4686, the high-ionization zone \te\ and \den\sii\ to calculate $\nu$. This exercise is solely for the inference of ICF(K), and more thorough measurements of He$^+$ are discussed in Paper IV.

All ten nebulae have $\nu$ $\leq$ 0.02, where the average of the sample is $\nu$ $\sim$ 0.008. The functional form of the \citet{amay2020} ICF is only calibrated on photoionization models with $\nu$ $\geq$ 0.02, such that the application of this ICF on the present sample represents a large systematic uncertainty. Nevertheless, to examine the general abundance trends of K in the low-metallicity ISM we use the estimated $\nu$ and the resulting ICF(K) to determine the relative K/Ar abundance. Recognizing that these ICFs are poorly constrained in this parameter space, we double the recommended uncertainty in the ICF from 0.05 dex to 0.10 dex. Similar to Cl and other trace elements, future calibrations of this ICF based on photoionization models with stellar ionizing sources and low-metallicity abundance patterns will provide more reliable estimates of K/Ar and K/H. Ionic abundances, ICFs, and total abundances measured for the LBT \yp\ Project nebulae are reported in Table \ref{t:abun}.

\section{Abundance Patterns in the Low-Metallicity ISM}\label{s:abun}

We now discuss the gas-phase chemical abundance patterns measured in the LBT \yp\ Project. As discussed in Paper I, this sample was originally selected based on the low gas-phase O/H reported in many previous works. Other targets were observed to provide a larger parameter space for the $Y$ vs.\ O/H relation. In Paper II, we showed that the sample spans 7.00 $\leq$ 12+log(O/H) $\leq$ 8.30 with an average uncertainty of $\sim$4\%. The LBT \yp\ Project observations, therefore, span an order of magnitude in O/H and present an opportunity to assess the evolution of various elements beyond He and O. In the following subsections, we explore the relative chemical abundance patterns in the metal-deficient ISM.

\subsection{Nitrogen Abundances}\label{s:noabun}

Of the elements observable in the rest-optical/NIR, N shows the strongest evolution with the O abundance in the ISM. The abundance patterns of N and O in the ISM have been the focus of numerous observational \citep{garn1990,vanZ2006,nich2012,belf2017,berg2020} and theoretical studies \citep{henr2000,vinc2016,vinc2018,johnson2023}. From these works, the N/O-O/H relation shows a roughly bimodal behavior. At low O/H abundance, the log(N/O) ratio remains constant at $\sim$ $-$1.4$\pm$0.1 \citep[e.g.,][]{berg2019}. This lack of evolution with O suggests a common nucleosynthetic origin of N and O at low-metallicities, often attributed to metallicity-insensitive N enrichment from massive stars. As the O abundance of the ISM increases, later stellar populations will have more seed C and O available to partake in the CNO cycle. The ``bottleneck" in this process is the reaction involving $^{14}$N($p$,$\gamma$)$^{15}$O \citep{henr2000}, such that $^{14}$N can build up in the H-burning zone while decreasing both C and O. If this N-rich material is liberated from the star (e.g., through metal-rich ejecta of asymptotic giant branch, AGB, stars), then more N is synthesized relative to O than the metallicity-insensitive N enrichment from either intermediate-mass or massive stars \citep[see Figure 2 in][]{vinc2016}. The net trend observed in local star-forming galaxies, therefore, is a combination of the primary, metallicity-insensitive N/O plateau at low-O/H and increasing N/O with increasing O/H at 12+log(O/H) $\gtrsim$8.0 \citep[e.g.,][]{pily2012}.

\begin{figure}[t]
   \centering
   \includegraphics[width=0.47\textwidth]{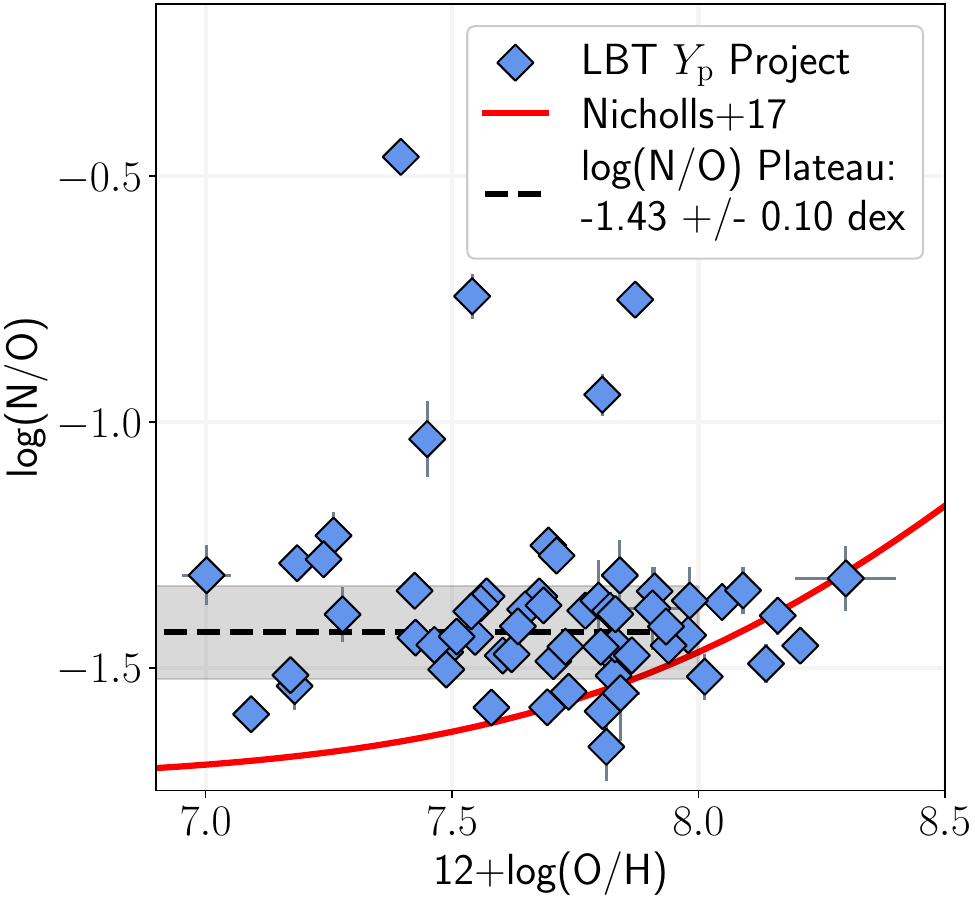}
   \caption{N/O vs.\ O/H measured from the LBT \yp\ Project sample. Consistent with prior findings in low-metallicity nebulae, the majority of the sample shows a constant N/O ratio. The average N/O measured in nebulae with 12+log(O/H) $<$ 8.0 and log(N/O) $<$ $-1.1$ is provided as a dashed black line and shaded area. The \citet{nich2017} N/O-O/H relation is plotted in red.}
   \label{fig:lno}
\end{figure}

Figure \ref{fig:lno} plots the N/O-O/H abundances measured in the LBT \yp\ Project sample, where the N/O-O/H trend observed in Galactic stars and \hii\ regions from \citet{nich2017} is plotted in solid red. Of the 62 nebulae, seven have 12+log(O/H) $>$ 8.0, the metallicity at which secondary N enrichment is expected to become significant \citep{henr2000}. Another five galaxies have log(N/O) $>$ $-$1.1, significantly enhanced relative to the remainder of the sample. We will discuss these intriguing objects in \S\ref{s:highno}. Excluding these twelve nebulae, the weighted average of the remaining N/O abundances is log(N/O)$_{Prim}$ $=$ $-$1.43 with standard deviation $\pm$0.10 (black dashed line and shaded area in Figure \ref{fig:lno}). This result is in good consistency with observations of similarly metal-poor nebulae in the local Universe \citep{vanZ2006,berg2019}.

\begin{figure*}[t]
   \centering
   \includegraphics[width=0.67\textwidth]{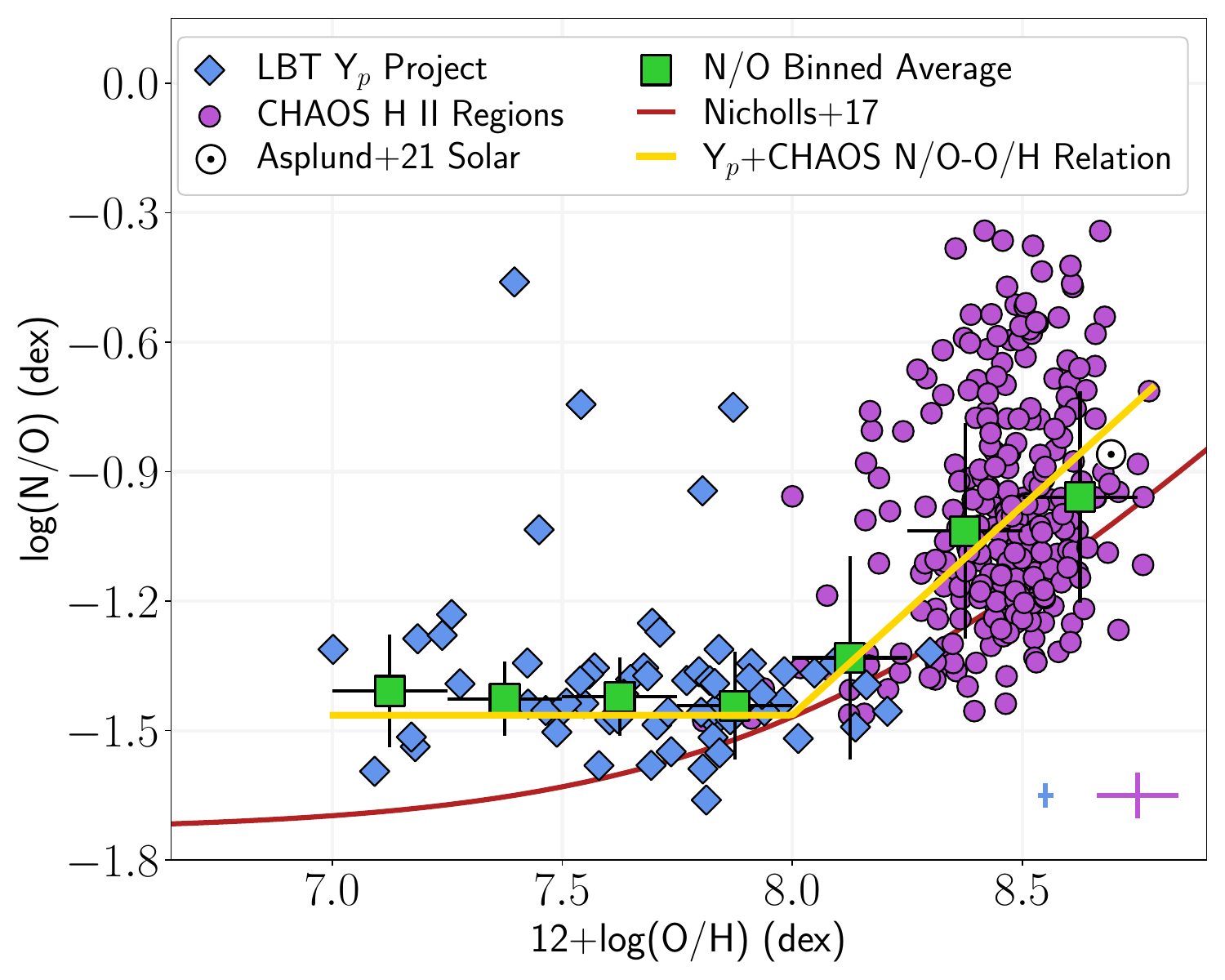}
   \caption{N/O vs.\ O/H measured in the LBT \yp\ Project nebulae (blue diamonds) and relatively metal-rich CHAOS \hii\ regions (purple circles). Median uncertainties from both samples are plotted in the bottom right. The total sample comprises 329 sources (267 extragalactic \hii\ regions and 62 metal-poor nebulae) with N/O and O/H abundances measured from LBT/MODS. Average N/O in O/H bins of 0.25 dex are plotted as green squares. The red line is derived from \citet{nich2017} using abundance trends in Galactic stars and \hii\ regions, while the solid gold line is an ODR fit to the LBT \yp\ Project and CHAOS N/O-O/H abundances using a piecewise function (see Equation \ref{eq:no2}). The binned averages and best-fit relation show a nearly constant N/O at 12+log(O/H) $<$ 8.2, then N/O rapidly increases owing to the contribution from metallicity-sensitive enrichment sources like AGB winds.}
   \label{fig:lno_full}
\end{figure*}

While the focus of this section is on the metal abundance patterns at low total metallicity, it is not obvious that there is a secondary N enrichment component from the LBT \yp\ Project observations alone. A sample of metal-rich nebulae is required to illustrate the effect of secondary N enrichment on the N/O abundance ratio. The LBT \yp\ Project observed few metal-rich nebulae by design, so we draw upon observations of extragalactic \hii\ regions acquired from the CHemical Abundances Of Spirals \citep[CHAOS,][]{berg2015} program. The benefit of this comparison is that both CHAOS and the LBT \yp\ Project utilize LBT/MODS for spectroscopy, such that the spectral resolution and wavelength coverage of the optical spectra are consistent\footnote{Note, however, that the CHAOS spectra have been resampled to 1 \AA\ per pixel. This does not impact the chemical abundance measurements, but does result in blended emission lines that limits the ability to measure \den\ from ions such as O$^+$ and Ar$^{3+}$.}. For this comparison, we use the \hii\ region observations in six spiral galaxies from \citet{berg2020}, \citet{roge2021}, and \citet{roge2022}. The \te\ and chemical abundances of all CHAOS \hii\ regions have been recomputed following the methodology outlined in \citet{roge2022}.

The resulting N/O-O/H abundances are plotted in Figure \ref{fig:lno_full}, where the 267 CHAOS \hii\ regions are represented as purple circles. As expected, the more metal-rich \hii\ regions show a significant evolution in N/O as metallicity increases; this was originally reported in \citet{berg2020} for four CHAOS galaxies, but the addition of $\sim$90 \hii\ regions in NGC~2403 and M~33 further illustrates that another enrichment mechanism is increasing the N abundance relative to the constant N/O plateau observed in the LBT \yp\ Project sample. The green squares represent the average N/O relative abundance in bins of 0.25 dex in O/H. From the averages, the N/O abundances at low O/H show a mild evolution from 12+log(O/H) $\sim$ 7.0 to 8.2 before increasing significantly toward the solar abundance ratio. From the 324 relative abundances (excluding the five N/O enhanced LBT \yp\ galaxies), we fit a piecewise function, assuming log(N/O) is constant at 12+log(O/H) $<$ 8.0 and increases linearly with log(O/H) above 8.0. Using \textsc{scipy} \textsc{ODR} to perform the fit considering the errors in both log(N/O) and 12+log(O/H), we find
\begin{equation}\label{eq:no2}
    \mbox{log(N/O)} = 
    \begin{cases}
        -1.46, & x < 8.0 \\
        -9.22 + 0.96\times x, & x \geq 8.0
    \end{cases}
\end{equation}
where $x$$=$12+log(O/H). This relation is plotted as a solid gold line in Figure \ref{fig:lno_full}, which reasonably fits the binned average N/O and O/H abundances in the star-forming nebulae.

We also fit the N/O-O/H relation using the same functional form as \citet{nich2017}:
\begin{equation}
    \mbox{log(N/O)} = \mbox{log(10}^{a} + \mbox{10}^{\mathrm{12+log(O/H)}-b}),
\end{equation}
where $a$ represents the plateau in N/O at low metallicity and $b$ sets the contribution from secondary N enrichment. The relation measured from the CHAOS and LBT \yp\ Project nebulae is
\begin{equation}\label{eq:no}
    \mbox{log(N/O)} = \mbox{log(10}^{-1.57} + \mbox{10}^{\mathrm{12+log(O/H)}-9.70}),
\end{equation}
implying the lowest inferred N/O ratio from metallicity-insensitive enrichment mechanisms is log(N/O) $=$ $-$1.57. We caution against the extrapolation of this simple expression to O/H that are lower than observed in most star-forming galaxies. Regardless, the two functional forms of the N/O-O/H relation have similar reduced $\chi^2$ statistics (21.9 and 22.5 for Eq.\ \ref{eq:no2} and \ref{eq:no}, respectively), and both reproduce the solar N/O ratio at 12+log(O/H) $=$ 8.69. There is large dispersion in the high-metallicity \hii\ regions that may depend on the different SFHs of the CHAOS galaxies \citep[see][]{berg2020}; overall, the relations from the combined LBT \yp\ Project and CHAOS \hii\ regions reasonably characterizes the combination primary and secondary N production and the average N/O-O/H abundance patterns in local nebulae.

\subsection{Neon}

Ne, S, and Ar abundances are frequently measured from optical spectra of low-metallicity star-forming galaxies. These elements are produced in massive stars along with O through the $\alpha$ process and released to the ISM via stellar winds and Core Collapse SNe (CCSNe); the relative abundance of $\alpha$/O is expected to remain constant if massive stars are the main enrichment source of the $\alpha$ process elements.
The abundances of Ne, S, and Ar have been observed to scale with O in local nebulae, and the $\alpha$/O ratios are in fairly good agreement with the solar abundance patterns \citep[e.g.,][]{este2025}. However, there may be other enrichment sources capable of producing the different $\alpha$ elements. For example, recent nucleosynthetic yield models predict that S and Ar are also produced in Type Ia SNe \citep{koba2020a}, such that the abundance of these elements are sensitive to the time delay between CCSNe and Type Ia SNe \citep[e.g.,][]{arna2022,koba2023,roge2024,rogers2026cecilia,bhattacharya2026}. Additionally, a metallicity-sensitive enrichment mechanism for Ne may be present in both massive stars \citep{woos1995} and AGB stars \citep{koba2011,koba2025}, which would introduce a correlation between Ne/O and O/H at high metallicity.

\begin{figure}[t]
   \centering
   \includegraphics[width=0.47\textwidth]{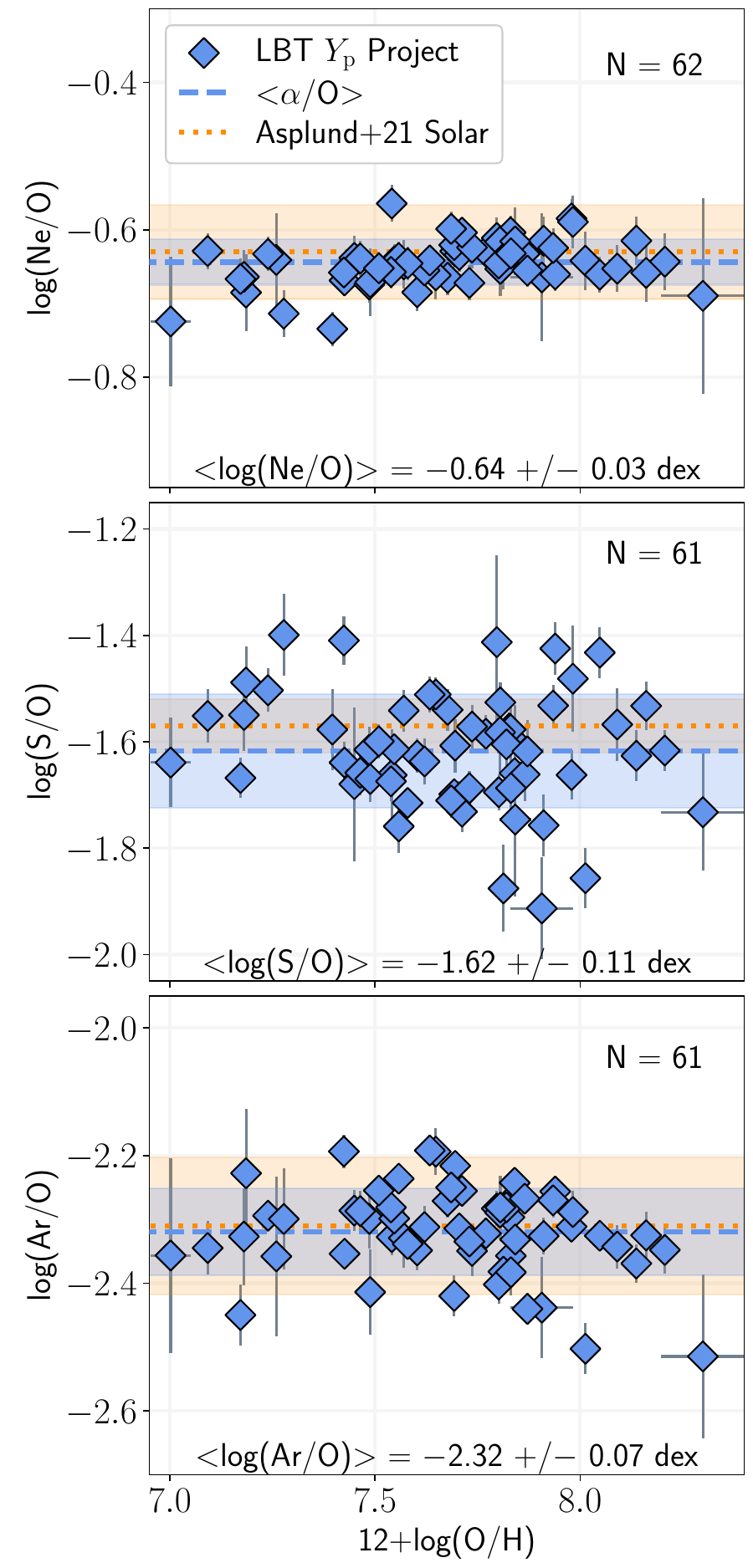}
   \caption{$\alpha$/O vs.\ O/H abundances for the LBT \yp\ Project nebulae. The number of Ne/O (Top), S/O (Middle), and Ar/O (Bottom) abundances are provided in the top right of each panel. The average $\alpha$/O ratio and its uncertainty (provided in the bottom of each panel) are represented as a dashed blue line and shaded area, respectively, while the solar value is shown as an orange dotted line. The relative abundances of the $\alpha$ elements show no trend with metallicity, and all agree with the solar ratios. This is related to a common nucleosynthetic origin for these elements (massive stars and CCSNe), although S/O and Ar/O are also sensitive to Type Ia enrichment.}
   \label{fig:pure_alphao}
\end{figure}

We plot the abundance ratios of Ne/O, S/O, and Ar/O vs.\ 12+log(O/H) for the LBT \yp\ Project sample in Figure \ref{fig:pure_alphao}. The dashed blue line and blue shaded region in each panel represents the average $\alpha$/O abundance ratio and the standard deviation measured from the sample, respectively, also provided in the bottom of each panel. The solar abundance ratios are plotted as an orange dotted line and shaded area. From this large, homogeneous sample of local, metal-poor galaxies, it is clear that there is little evolution of $\alpha$/O with O/H, consistent with prior studies \citep{thua1995,izot2006,berg2020,roge2021,roge2022,arel2024}.

Of these relative abundances, the gas-phase Ne/O shows the smallest dispersion at just 0.03 dex, and the average Ne/O agrees very well with the solar abundance ratio.
Using ODR to fit a linear relation to the log(Ne/O) vs.\ 12+log(O/H) trend, we find a slope of 0.043$\pm$0.014. The Pearson's correlation coefficient for these data is $r$$=$0.35 with a $p$ value of 0.005, suggesting that the slight increase in Ne/O with O/H is significant. A marginal correlation between Ne/O and O/H has been observed in other large samples of extragalactic nebulae. For example, \citet{izot2006} report a significant evolution of log(Ne/O) with 12+log(O/H) in over 400 star-forming galaxies in SDSS, finding a slope of 0.088$\pm$0.018. This trend was used to argue for O depletion onto dust grains in metal-rich environments, thereby producing elevated Ne/O at high O/H. \citet{este2025} determined the Ne/O abundances in 671 star-forming galaxies \citep[including those from][]{izot2006} from the DESIRED-E sample. The log(Ne/O) vs.\ 12+log(O/H) slope was determined to be 0.051$\pm$0.006, consistent with the value measured in the LBT \yp\ Project nebulae. Finally, \citet{scholte2026} find a strong correlation between Ne/O-O/H from $\sim$48000 galaxies observed as part of the DESI survey, with a slope of 0.097$\pm$0.002. 

It is important to note that these larger surveys consider more metal-rich star-forming galaxies. There are only seven galaxies in the LBT \yp\ Project with 12+log(O/H) $>$ 8.0, but this metallicity range describes the majority of the nebulae analyzed in \citet{izot2006}, \citet{este2025}, and \citet{scholte2026}. For Ne/O measured in low metallicity galaxies, \citet{arel2024} find that the CLASSY galaxies exhibit a constant Ne/O ratio with an average log(Ne/O) $=$ $-$0.63$\pm$0.06. The CLASSY galaxies with 12+log(O/H) $\gtrsim$ 8.5, however, show large scatter to super-solar Ne/O. Further, some metal-rich extragalactic \hii\ regions have been observed to have sub-solar Ne/O \citep[e.g.,][]{kenn2003b,crox2016,berg2020,roge2022}, further complicating the interpretation of Ne enrichment in the ISM.

The relatively shallow Ne/O-O/H slope observed in the present sample indicates that there is little evolution in Ne/O at low metallicities. This trend could be a consequence of negligible O dust depletion at low O/H and a lack of enrichment from a potential metallicity-sensitive mechanism of Ne production. Increased Ne abundance or inflated scatter in Ne/O at high metallicities could be the result of a poorly constrained ICF in low-ionization environments. The majority of Ne in the high-ionization ISM is in the Ne$^{2+}$ state, such that the ICF is accounting for a fractionally smaller amount of missing Ne$^+$. In high-metallicity environments with low ionization conditions, the correction for Ne$^+$ accounts for the majority of the total Ne abundance. Therefore, the shape of the ICF fundamentally sets the Ne/O abundance trends; even modern ICFs like those of \citet{amay2021} produce large scatter in Ne/O when the ionization conditions in the gas are low, as has been shown in the CHAOS \hii\ regions with ionic fractions O$^{2+}$/O $<$ 0.2 \citep[see discussion in][]{roge2022}. Considering the minimal dust depletion and systematics related to the ICF, the robust Ne/O abundances in the LBT \yp\ Project support a lock-step evolution of Ne and O in the ISM that produces a ratio consistent with the solar value, as expected for two elements produced primarily via the $\alpha$ process in massive stars.

Observations of the \neiii\ strong lines in low-metallicity nebulae, such as those of the LBT \yp\ Project, provide a fairly robust measurement of the majority of Ne in the gas. Indeed, direct constraints on the ISM Ne/O abundance in high-ionization, metal-poor distant galaxies observed with JWST have found that the gas-phase Ne/O generally agrees with the solar abundance ratio \citep[e.g.,][]{arel2022,curt2024,marq2024,stan2025,welc2025}. The lock-step evolution of Ne and O makes the optical \neiii\ strong lines an enticing, indirect tracer gas-phase metallicity \citep[e.g.,][]{scholte2025,sanders2026}. While these strong line diagnostics have the same systematic limitations as other optical diagnostics \citep[e.g., sensitivity to ionization parameter, \te, \den, and chemical abundance ratios, see][]{pagel1980,skil1989,kewl2002,pere2009}, the use of rest-frame blue optical lines like \oii$\lambda$3727 and \neiii$\lambda$3869 are useful when \oiii$\lambda$4959,5007 redshift out of JWST/NIRSpec's wavelength coverage.

For local nebulae, metallicity diagnostics utilizing the mid-IR (MIR) [\ion{Ne}{2}] and \neiii\ fine-structure lines offer a promising alternative. These MIR diagnostics show strong correlations with O/H due to the constant Ne/O ratio in the ISM, are insensitive to dust attenuation, and relatively robust to ionization, \te, and \den\ biases compared to their optical counterparts \citep[see][]{fern2021,rogers2026ne23}. In summary, the constant Ne/O ratio in the low-metallicity ISM has useful implications for the ability to infer 12+log(O/H) from an emission line spectrum, either from rest-optical observations of metal-poor high-\emph{z} galaxies or MIR spectra of local nebulae.

\subsection{Sulfur and Argon}

The middle panel of Figure \ref{fig:pure_alphao} plots log(S/O) vs.\ 12+log(O/H). The average log(S/O) measured in the LBT \yp\ Project sample is $-$1.62$\pm$0.11 with no significant evidence for a correlation with O/H 
(slope of $-$0.04$\pm$0.05, Pearson $r$ $=$ $-$0.16, $p$ $=$ 0.20). The scatter in gas-phase S/O is larger than that of Ne/O and Ar/O and similar to the scatter observed in recent compilations of gas-phase S abundances in ionized nebulae \citep{arel2024,este2025}. Multiple systematic and observational uncertainties can influence the scatter in S/O.
First, the total S abundance is determined from the strong lines of S$^+$ and S$^{2+}$, where the missing abundance of S$^{3+}$ is accounted for with an ICF. Given the relatively high ionization conditions observed in the LBT \yp\ Project, the ICF must account for a large fraction of missing S that is unconstrained from the optical/NIR data, and the shape of the ICF can enhance the dispersion in S/O. Second, telluric absorption features can contaminate the NIR lines of \siii. While we attempt to mitigate this problem with the \siii\ line intensity ratios (see Paper II), this approach does not work if both \siii\ strong lines are contaminated or if the observed \siii$\lambda$9532 is redshifted beyond MODS's wavelength coverage. Finally, the S$^+$/H$^+$ can be fairly uncertain owing to an indirect estimate of the low-ionization zone \te, but also due to other biases of the \sii\ strong lines (e.g., \den\ inhomogeneities).

\begin{figure}[t]
   \centering
   \includegraphics[width=0.48\textwidth]{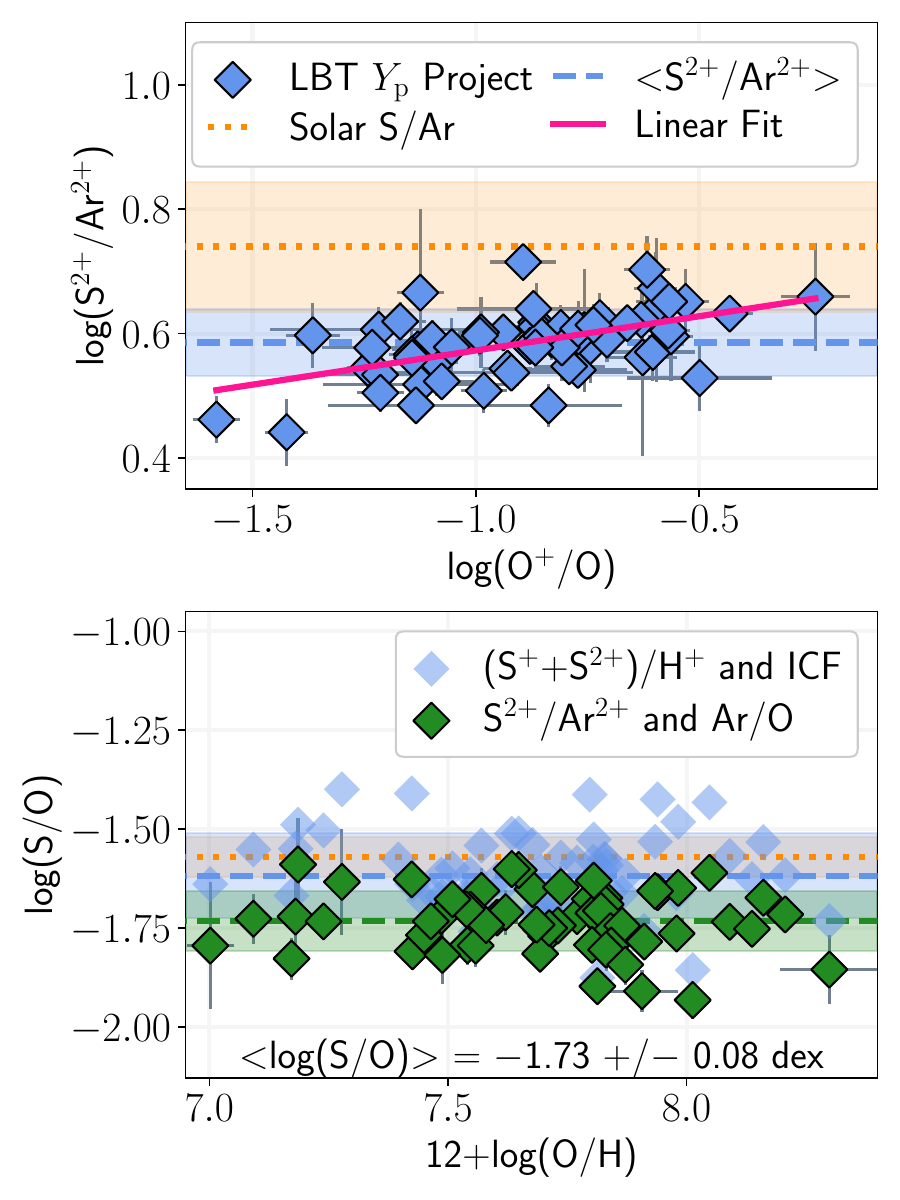}
   \caption{Top: log(S$^{2+}$/Ar$^{2+}$) vs.\ log(O$^{+}$/O) for the LBT \yp\ Project sample. The relative ionic abundance of S$^{2+}$/Ar$^{2+}$ is lower than the solar S/Ar abundance ratio (dotted orange line) and is correlated with O$^{+}$/O (solid pink line). Bottom: The inferred S/O abundances using log(S$^{2+}$/Ar$^{2+}$) $-$ log(Ar/O) (green) and the reported S/O abundances from Figure \ref{fig:pure_alphao} (blue). The scatter in S/O is lower than reported in Figure \ref{fig:pure_alphao}, indicating that some of the dispersion in S/O may be related to the choice of ICF. However, the average S/O (dashed green) is now more than 0.1 dex below the solar S/O abundance.}
   \label{fig:s3ar3}
\end{figure}

We explore the source of this dispersion by comparing the relative ionic abundance of S$^{2+}$ and Ar$^{2+}$. Both ions occupy the intermediate-ionization zone with the same adopted \te. Previous studies have proposed using S$^{2+}$/Ar$^{2+}$ as an ICF-insensitive tracer of the relative S/Ar abundance \citep{stevenson1993,kenn2003b,crox2016}. The top panel of Figure \ref{fig:s3ar3} plots log(S$^{2+}$/Ar$^{2+}$) against log(O$^+$/O). The average relative abundance is log(S$^{2+}$/Ar$^{2+}$) $=$ 0.59$\pm$0.05 (dashed blue line), which is in good agreement with the average log(S$^{2+}$/Ar$^{2+}$) ionic abundance ratio reported in the low-metallicity CLASSY galaxies \citep{arel2024}. However, this is more than 0.1 dex below the solar S/Ar ratio, plotted as the dotted orange line. Furthermore, a significant correlation between S$^{2+}$/Ar$^{2+}$ and O$^+$/O is measured. The linear relation plotted in pink is log(S$^{2+}$/Ar$^{2+}$) $=$ 0.68 $+$ 0.11$\times$log(O$^+$/O) (Pearson $r$ $=$ 0.47, $p$ $=$ 1.6$\times$10$^{-4}$). Despite the similarity in IP of the two ions, it is clear that S$^{2+}$/Ar$^{2+}$ is dependent on the ionization state of the gas at log(O$^+$/O) $\lesssim$ $-$0.5, or O$^+$/O $<$ 0.32.

Given the lower scatter in Ar/O abundances and S$^{2+}$/Ar$^{2+}$ compared to S/O, we estimate the log(S/O) as log(S$^{2+}$/Ar$^{2+}$) $-$ log(Ar/O). This approach tests whether the scatter in S/O abundances is related to the ICF, where the lower scatter in Ar/O is the result of a decreased dependence on ICF(Ar) at high ionization. The resulting S/O abundances are plotted as green points in the bottom panel of Figure \ref{fig:s3ar3}, where the light blue points are the S/O abundances measured using the \sii\ and \siii\ lines and the \citet{izot2006} ICF. We observe a 0.03 dex reduction in scatter compared to the standard deviation in S/O abundances in Figure \ref{fig:pure_alphao} (0.11 dex), demonstrating that ICF(S) may contribute to the dispersion observed in S/O in addition to observational effects. The average is now inconsistent with solar S/O ratio owing to the offset of the S$^{2+}$/Ar$^{2+}$ abundances from the solar ratio.

This exercise calls into question the assumption that S$^{2+}$/Ar$^{2+}$ $\approx$ S/Ar, which implies the relative abundance of S/Ar in the low-metallicity ISM is, on average, more than 0.1 dex sub-solar. In contrast, high-metallicity \hii\ regions have been observed to have S$^{2+}$/Ar$^{2+}$ in excess of the solar S/Ar abundance \citep[particularly at O$^{+}$/O $>$ 0.6, see][]{crox2016,berg2020}. In combination, these trends would suggest a metallicity-dependent evolution of S/Ar. However, this is unexpected from nucleosynthesis yields of S and Ar: both CCSNe and Type Ia SNe yields predict that the S/Ar ratio should remain relatively constant \citep{koba2020b}.

The offset in S$^{2+}$/Ar$^{2+}$ below the solar S/Ar ratio and the correlation with O$^+$/O in Figure \ref{fig:s3ar3} are likely reflections of the small difference in IP of S$^{2+}$ (23.3 eV) and Ar$^{2+}$ (27.6 eV). In the high-ionization ISM, a larger fraction of S will triply ionize compared to Ar.
There may be an ionization range where S$^{2+}$/Ar$^{2+}$ $\approx$ S/Ar is valid, particularly when the majority of S and Ar are in the doubly ionized states. For high-ionization objects, such as the LBT \yp\ Project nebulae and many high-\emph{z} star-forming galaxies, we caution against the use of S$^{2+}$/Ar$^{2+}$ to infer S/Ar or S/O abundances.

The bottom panel of Figure \ref{fig:pure_alphao} reveals that the average Ar/O abundance observed in the LBT \yp\ Project nebulae agrees with the solar abundance Ar/O ratio. The dispersion in Ar/O is 0.07 dex, and there are some nebulae with significantly sub-solar Ar/O. \citet{arel2024} observed a similar trend for CLASSY galaxies without a direct constraint on the Ar$^{3+}$/H$^+$ abundance, indicating this dispersion may also be related to the Ar ICF. Of the 9 galaxies with Ar/O 1$\sigma$ below the average Ar/O ratio, three have only one direct ionic abundance of Ar measured. The remaining six have direct measurements of Ar$^{2+}$ and Ar$^{3+}$ abundances, suggesting that a small fraction of the sample is characterized by marginally Ar-deficient ISM.

The gas-phase abundance of S and Ar has been the subject of renewed interest owing to the enrichment sources of the two elements. As noted previously, nucleosynthesis yields predict that Type Ia SNe can produce S and Ar \citep[e.g.,][]{koba2020a}. This implies that the S/O and Ar/O abundance ratios are marginally sensitive to the SFH, where Type Ia SNe can produce additional S and Ar without significant O enrichment \citep[after a significant time delay,][]{koba2009}. Much of the attention has been on the Ar/O abundance ratios, since the ICF(Ar) is minimal for nebulae with direct Ar$^{2+}$ and Ar$^{3+}$ abundances and the \ariii$\lambda$7135 emission line can be detected with JWST/NIRSpec up to redshifts of \emph{z} $\sim$ 6.4. \citet{arna2022} first noted an evolution of Ar/O in a sample of PNe in M31, where the distinct populations of older and recently-formed PNe were used to retrace the SFH of M31 \citep[see also][]{koba2023}.

Compared to the M31 PNe, other local star-forming galaxies show a mild evolution in Ar/O vs.\ O/H or Ar/H. \citet{bhattacharya2026} found a marginal mass-dependent evolution of Ar/O in SDSS galaxies, which may be related to differences in SFH and star-formation efficiency (SFE) between low- and high-mass galaxies.
\citet{este2025} compare the Ar/O abundances in the DESIRED-E sample to the Galactic chemical evolution models, where the empirical data suggest a relatively flat Ar/O trend compared to the models.
Nevertheless, there are low-metallicity nebulae in the DESIRED-E sample with significantly sub-solar Ar/O, extending to log(Ar/O) $\sim$ $-$2.60 at 12+log(O/H) $<$ 8.0. The average Ar/O ratio observed in the LBT \yp\ Project sample is in good agreement with the DESIRED-E sample and other samples of metal-poor nebulae \citep[][]{arel2024}, and the relatively small scatter prevents a definitive assessment of the potential enrichment mechanisms of Ar in individual galaxies.

If S/O and Ar/O are sensitive to enrichment from Type Ia SNe, then high-\emph{z} star-forming galaxies with abbreviated SFHs should provide the best opportunity to observe the evolution of these abundance ratios. However, the number of high-\emph{z} galaxies with comparably high-quality spectra and multiple auroral line detections remains small. For example, the direct abundances patterns in Q2343-D40, a galaxy at \emph{z}$\sim$3 observed by the CECILIA survey \citep{stro2023}, are constrained from directly-measured \te\oiii\ and \te\siii. From these temperatures, \citet{roge2024} report that the ISM in Q2343-D40 has log(Ar/O)$=$$-$2.80$\pm$0.12, lower than observed in any of the LBT \yp\ Project nebulae. Sub-solar Ar/O in the ISM has been reported in the other samples of \emph{z} $>$ 2 galaxies, including the remaining $\sim$20 galaxies in CECILIA \citep{rogers2026cecilia}, nine galaxies from EXCELS \citep[][]{stan2025}, and from the strong-line trends of 43 galaxies in the AURORA survey \citep{foley2026}. These findings suggest that the abundance of Ar is sensitive to the SFH, where the sub-solar Ar/O in typical high-\emph{z} galaxies is a consequence of minimal Type Ia enrichment.
Not all high-\emph{z} galaxies have sub-solar Ar/O abundance ratios \citep[see][]{isob2023,welc2025,mori2025}, which may be related to differences in the SFHs of distant galaxies.

It should be emphasized that most chemical abundance studies of high-\emph{z} galaxies lack a direct measurement of \te\siii\ or the \ariv\ CELs, which may introduce systematic biases owing to the use of \te\ scaling relations and ICFs, respectively. Similarly, S remains under-studied in distant galaxies compared to Ar owing to the required wavelength coverage. Future JWST observations will reveal whether there is a redshift evolution of abundance ratios like S/O and Ar/O, thereby elucidating the different enrichment mechanisms that drive their production.

Nevertheless, the average Ar/O and S/O abundances of high-\emph{z} star-forming galaxies are in contrast to those observed in the metal-poor nebulae of the LBT \yp\ Project. Again, this disagreement is likely driven by differences in the SFHs of the high-\emph{z} and local metal-poor galaxies.
Indeed, studies of resolved stellar populations in local, low-mass, metal-poor dwarf galaxies indicate that these systems can have highly stochastic SFHs \citep{weisz2014} or may briefly quench following cosmic reionization \citep[e.g.,][]{mcquinn2024}, permitting the buildup of Type Ia products such as S and Ar over longer epochs or during periods of quiescence. The rising and abbreviated SFHs at \emph{z} $>$ 2 should produce a strong difference in the enrichment from CC and Type Ia SNe, which could be traced with gas-phase abundances like S/O and Ar/O.
In summary, the S/O and Ar/O abundances observed in the LBT \yp\ Project nebulae and other local galaxies are consistent with the solar ratio, implying more episodic or prolonged SFHs compared to high-\emph{z} galaxies.

\subsection{Odd-Z Elements}\label{s:oddz}

We now examine the relative abundance trends of Cl and K, which are adjacent to the $\alpha$ process elements S, Ar, and Ca on the periodic table. The production of Cl and K in massive stars is expected to be regulated by the available $\alpha$ process elements in the core of the massive star as it undergoes O burning \citep{woos1995}. The main reactions that produce Cl are $^{34}$S($p$,$\gamma$)$^{35}$Cl and $^{36}$Ar($n$,$\gamma$)$^{37}$Ar followed by the decay of radioactive $^{37}$Ar to $^{37}$Cl \citep{clay2003}. Similarly, K is synthesized during O burning through the reactions $^{38}$Ar($p$,$\gamma$)$^{39}$K and $^{40}$Ca($n$,$\gamma$)$^{41}$Ca, where the half-life of $^{41}$Ca is long ($\sim$10$^5$ yr) and can persist after the supernova explosion \citep{clay2003}. Since these reactions take place during O burning and require a significant abundance of $\alpha$ process elements, the abundance of Cl and K may behave similarly to the other $\alpha$ elements in the ISM.

\subsubsection{Chlorine}

\begin{figure}[t]
   \centering
   \includegraphics[width=0.47\textwidth]{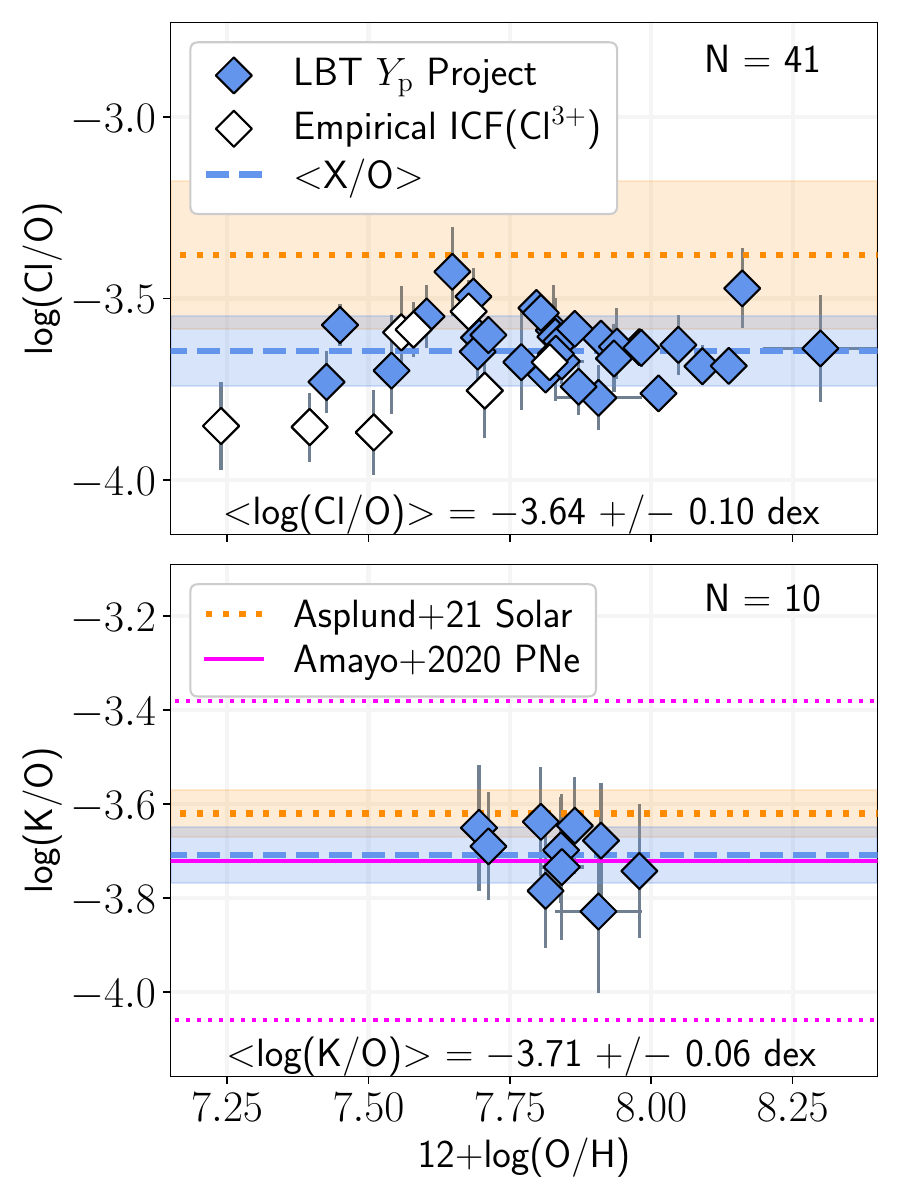}
   \caption{Abundance trends for Cl/O (Top) and K/O (Bottom) vs.\ 12+log(O/H). The average and solar abundance ratios follow the same notation as Figure \ref{fig:pure_alphao}, and Cl abundances derived using Equation \ref{eq:icfcl3} are plotted as white diamonds. We also include the average K/O measured in metal-rich PNe from \citet[][]{amay2020} as a solid pink line (standard deviation in dotted pink). The average Cl/O and K/O abundance ratios in the low-metallicity ISM are marginally sub-solar. There is no trend in Cl/O or K/O with O/H, similar to the behavior of typical $\alpha$ process elements like Ne. This suggests that the production of these odd-Z elements in low-metallicity nebulae is similar to that of the $\alpha$ elements.}
   \label{fig:adj_alpha}
\end{figure}

The top panel of Figure \ref{fig:adj_alpha} plots log(Cl/O) vs.\ 12+log(O/H) in the LBT \yp\ Project nebulae. We highlight the eight nebulae with Cl measurements using the empirical ICF of Equation \ref{eq:icfcl3} as white diamonds. The average Cl/O and solar ratio are plotted as a blue dashed line and orange dotted line, respectively, where the metal-poor nebulae have Cl/O offset 0.26 dex below the solar ratio. The scatter in Cl/O for the 41 nebulae is similar to that of S/O, and no evolution with total O abundance is observed (Pearson $p$ $=$ 0.16). Prior studies of more metal-rich \hii\ regions, including those in the Milky Way \citep{este2015,arel2020M} and M~33 \citep{roge2022}, found Cl/O abundances in good agreement with the solar ratio and with negligible evolution as a function of 12+log(O/H) or galactocentric radius. \citet{orte2025} reported similar results in the DESIRED \hii\ regions, although Cl/O in the low-metallicity galaxies in the DESIRED sample show scatter toward sub-solar abundances. \citet{arel2024} also find that the Cl/O abundances are sub-solar in 15 CLASSY galaxies, with an average log(Cl/O)$=$$-$3.62$\pm$0.17 that is in good agreement with the average in Figure \ref{fig:adj_alpha}.

Considering these results, it appears that the average log(Cl/O) abundance in the metal-poor ISM is $\sim$0.2 dex below the solar ratio, although the solar ratio is relatively uncertain. Measuring Cl abundance in stars is exceptionally challenging due to the low dissociation energy of the HCl molecule; therefore, stellar Cl abundance measurements require stars with low effective temperatures \citep[e.g.,][]{maas2016}. In the Sun, the Cl abundance is constrained through HCl absorption features in sunspot spectra \citep{hall1972}. \citet{orte2025} recommended using the recent meteoric measurement of 12+log(Cl/H) from CI-chondrites \citep[5.26$\pm$0.06,][]{lodders2023} for the solar Cl abundance, which translates to a solar value of log(Cl/O) $=$ $-$3.46$\pm$0.07. Even with this updated solar abundance, the average Cl/O abundance in the LBT \yp\ Project is sub-solar at greater than 1$\sigma$.

\subsubsection{Potassium}

Gas-phase K abundances provide further insight into the enrichment mechanisms for the odd-Z elements, although the available emission features in the optical/NIR are rarely detected. Thus, the ten nebulae plotted in Figure \ref{fig:adj_alpha} represent a first look at the K abundance patterns in the low-metallicity ISM. While the errors on the individual K/O abundances are relatively large and the 12+log(O/H) range is narrow, the log(K/O) ratios show no evolution with O/H and the average ($-$3.71$\pm$0.06) is marginally lower than the \citet{aspl2021} Solar ratio ($-$3.62$\pm$0.05). We also provide the average K/O abundance measured in 28 PNe from \citet{amay2020} in solid pink, where the dotted pink lines represent the standard deviation in the relative abundances. All the PNe are significantly more metal-rich than the ten nebulae in Figure \ref{fig:adj_alpha} and exhibit much larger scatter, which may be related to depletion of K onto dust grains \citep{amay2020}.

Unlike Cl, stellar K abundances are more accessible owing to \ion{K}{1} absorption features in the rest-NIR. K abundances have been measured in individual stars of globular clusters and used to assess different enrichment scenarios in their early formation \citep[e.g.,][]{cohe2012,pran2017,carr2022}. In Galactic stars, K behaves similarly to an $\alpha$-process element: at [Fe/H]$<$$-$1.0, [K/Fe] is generally $\sim$0.2-0.4 and marginally anti-correlated with [Fe/H], then [K/Fe] rapidly decreases to $\sim$0 as [Fe/H] increases beyond $-$1.0 owing to the onset of Type Ia SNe \citep{andr2010,zhao2016,regg2019,rein2024}.
However, most Galactic chemical evolution models severely under predict the Cl and K abundance at all metallicities (see Figure 21 in \citealt{koba2020b}, also see discussion in \citealt{rein2024}). 

The abundances of Cl and K in the LBT \yp\ Project nebulae, therefore, suggest that these elements scale with $\alpha$-process elements in the low-metallicity ISM. The average Cl/O and K/O are marginally below the abundance ratios observed in the Sun ($\sim$0.2 and 0.1 dex for Cl and K, respectively), which might suggest increased odd-Z element production at higher metallicities. An expanded sample of K abundances, from nebulae that span a larger range in metallicity and with appropriately-calibrated ICFs, would better constrain the chemical evolution of the odd-Z elements in conjunction with Cl/H. The slightly sub-solar abundance ratios are not as significant as the discrepancy observed in chemical evolution models \citep[$\sim$0.8 dex,][]{koba2020b}. It may be necessary to revise the nuclear reaction channels that facilitate odd-Z element production in these models, which may then be able to reproduce the more abundant Cl and K observed in the Sun and extragalactic \hii\ regions. 

\subsection{Iron}

\begin{figure}[t]
   \centering
   \includegraphics[width=0.48\textwidth]{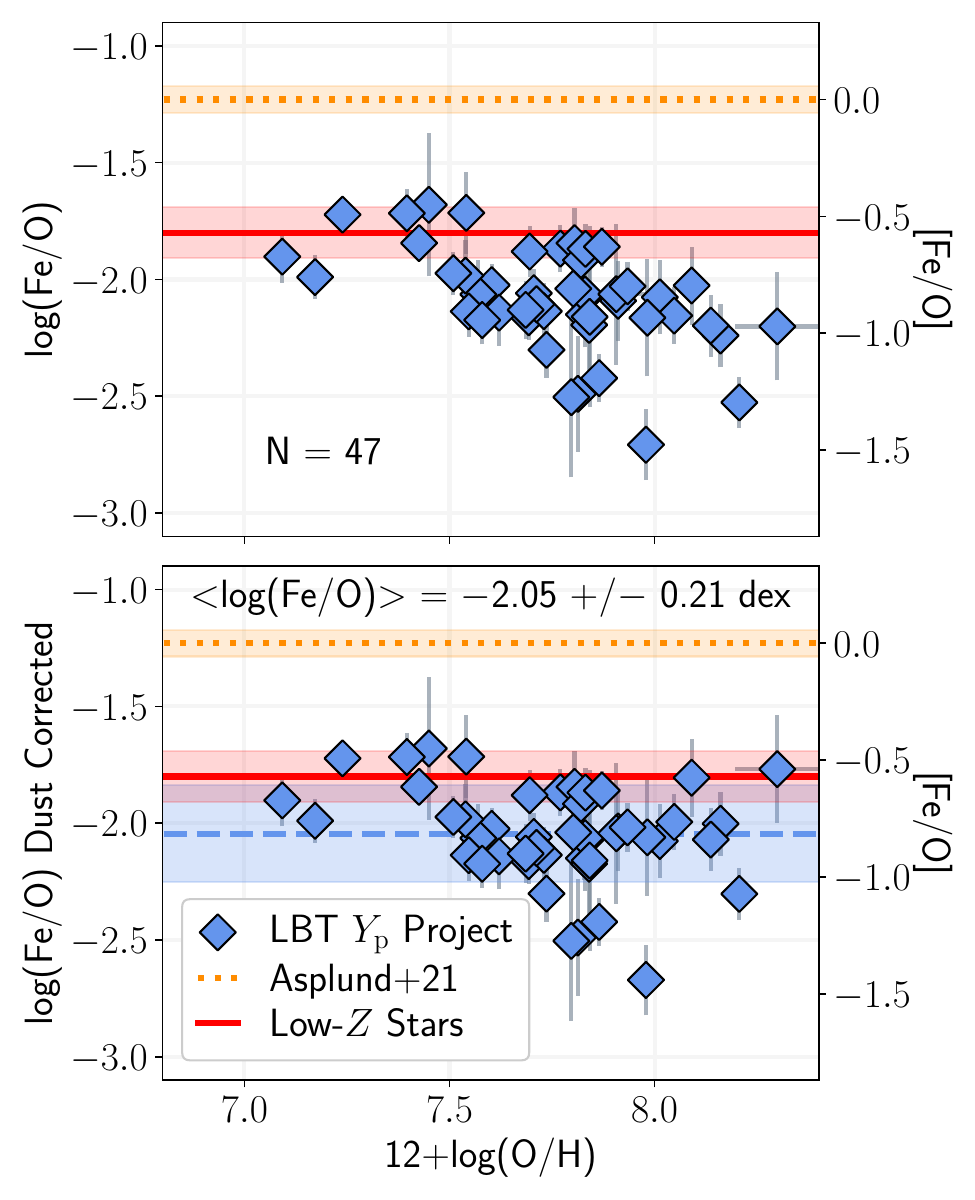}
   \caption{Top: log(Fe/O) vs.\ 12+log(O/H). The anti-correlation in Fe/O-O/H is related to dust depletion, an affect that is prominent even in the low-metallicity ISM. Bottom: Same as top figure, except Fe/O has been corrected for the fraction of Fe in dust grains as estimated using Eq 1 of \citet{mend2024}. The anti-correlation is no longer significant, and the average Fe/O abundance is in fairly good agreement with the Fe abundance trends in low-metallicity stars. The LBT \yp\ Project nebulae have significantly sub-solar Fe/O (average in blue dashed line). If the ionizing stars have a similar abundance pattern, then they would contribute to the high ionization conditions present in the sample.}
   \label{fig:feo}
\end{figure}

The final elemental abundance measured in the LBT \yp\ Project nebulae is Fe. The top panel of Figure \ref{fig:feo} plots the relative abundance of Fe/O in 47 low-metallicity nebulae. The relative abundance of Fe/O in metal-poor Galactic stars \citep[compiled from the literature by][see references therein]{mend2024} is plotted as a solid red line. The trend observed in the top panel is one of decreasing gas-phase Fe/O as O/H increases, which has been observed in other samples of local star-forming galaxies \citep{rodr2002,rodr2005,izot2006}. The anti-correlation between Fe/O and O/H is understood in terms of Fe depletion onto dust grains, a process that is most prevalent in metal-rich environments. The most metal-poor galaxies in the LBT \yp\ Project have Fe/O ratios that agree with the abundance patterns in low-metallicity stars: the average log(Fe/O) at 12+log(O/H) $<$ 7.60 is $-$1.91$\pm$0.17 compared to the stellar abundance ratio of log(Fe/O) $=$ $-$1.80$\pm$0.11. The consistency may be the result of minimal dust to promote Fe depletion at such low total metallicity \citep[see also][]{mend2024}. However, at 12+log(O/H) $>$ 7.60, or 8\% solar O abundance, the relative Fe/O abundance begins to noticeably decrease below the low-metallicity stellar abundance ratio. Without accounting for Fe depletion onto dust grains, it is challenging to interpret Fe/O abundance.

\citet{mend2024} examined the gas-phase Fe/O and Fe/N abundances in over 450 star-forming regions compiled as part of DESIRED. In addition to the anti-correlation between Fe/O and O/H, the DESIRED nebulae exhibit a similar decrease in Fe/N with increasing N/H. Arguing for similar timescales of N and Fe production from AGB stars and Type Ia SNe, respectively, \citet{mend2024} used the anti-correlation between Fe/N vs.\ N/H to estimate a dust correction for the gas-phase Fe abundances. Equation 1 in \citet{mend2024} is reported to be valid in the range 12+log(O/H) $\geq$ 7.6 and 12+log(N/H) $\geq$ 6.3; however, the Fe fraction in dust grains inferred from this equation can result in unphysical values (i.e., a negative fraction) if 12+log(N/H) $\lesssim$ 6.50. Further, the electron temperatures and the total O abundances reported for the DESIRED nebulae are corrected for biases introduced by temperature fluctuations, either using direct detections of the \ion{O}{2} recombination lines or the approach outlined \citet{mend2023N}. While we do not correct our O/H abundances for temperature fluctuations owing to a lack of faint recombination lines in the low-metallicity gas (see Paper II), we estimate the fraction of Fe depleted onto dust grains relative to the total Fe abundance using Eq 1 from \citet{mend2024}. This is only possible for the 13 LBT \yp\ Project nebulae satisfying 12+log(O/H) $\geq$ 7.6 and 12+log(N/H) $>$ 6.5, excluding the N-enhanced galaxies with log(N/O) $>$ $-$1.10.

We plot the dust-corrected Fe/O abundances in the bottom panel of Figure \ref{fig:feo}, where the anti-correlation between Fe/O and O/H is much weaker (Pearson $r$ $=$ $-$0.28, $p$ $=$ 0.06). The average gas-phase Fe/O (dashed blue line) is consistent with the Fe/O ratio in low-metallicity stars at the 1$\sigma$ level, albeit offset to lower Fe/O. Ultimately, the offset in Fe/O could be related to Fe dust depletion that remains unaccounted for, or shortcomings related the Fe ICF. As discussed in \S\ref{s:othericf}, the lack of Fe$^{3+}$ abundances results in large ICFs in the high-ionization ISM. Higher resolution spectroscopy may aid in this endeavor, as resolving the faint \feiv$\lambda$6740 emission line from the relatively strong \sii$\lambda$6731 is essential for a direct measurement of the Fe$^{3+}$/H$^+$. Alternatively, UV observations of the stellar continuum may aid in the inference of the Fe abundance of the ionizing stars, which may be related to the gas-phase Fe abundance.

Systematic uncertainties aside, the abundances in Figure \ref{fig:feo} suggest significantly sub-solar Fe/O in the low-metallicity ISM. Sub-solar gas-phase Fe/O has been measured in local extreme emission line galaxies \citep[e.g.,][]{berg2021}, and inferred from the rest-UV and optical spectra of high-\emph{z} galaxies \citep[e.g.,][]{stei2016,stro2018,cull2019,cull2021,stan2024}. If the stellar abundance patterns reflect the ISM metallicity at the time of star formation, this would indicate that the massive stars producing the \hii\ regions in the LBT \yp\ Project nebulae have low Fe abundances. This would reduce the impact of line blanketing in the stellar atmosphere, leading to harder ionizing spectra and high ionization conditions in the gas. Such conditions are clearly observed in the LBT \yp\ Project (Paper II), although direct measurements of the stellar abundances in these galaxies are challenging without UV spectroscopy. A combined UV and optical analysis requires a careful treatment of the systematic uncertainties in the Fe abundances and is beyond the scope of this paper. We note that the galaxies of the LBT \yp\ Project sample provide excellent laboratories for assessing the interplay between metal-poor stars and gas, similar to the average conditions that describe high-\emph{z} galaxies.

\section{Helium Abundance Trends}\label{s:heabun}

The primary goal of the LBT \yp\ Project is to provide a precision measurement of the primordial He abundance. As discussed in Section \S\ref{s:intro}, this has traditionally been achieved through measurements of He/H in low-metallicity gas and an extrapolation back to O/H $\sim$ 0 \citep{peimbert1974, peimbert1976}. Given the many elemental abundances measured from the MODS spectra, it is of interest to compare the He number density, y $=$ He/H, to these elements to look for secondary trends. The N/O relative abundance was investigated in Paper IV in the context of the scatter around the y vs.\ O/H relation, and we can now examine any correlation between y and other total and relative metal abundances.

\begin{figure*}[t]
   \centering
   \includegraphics[width=0.98\textwidth]{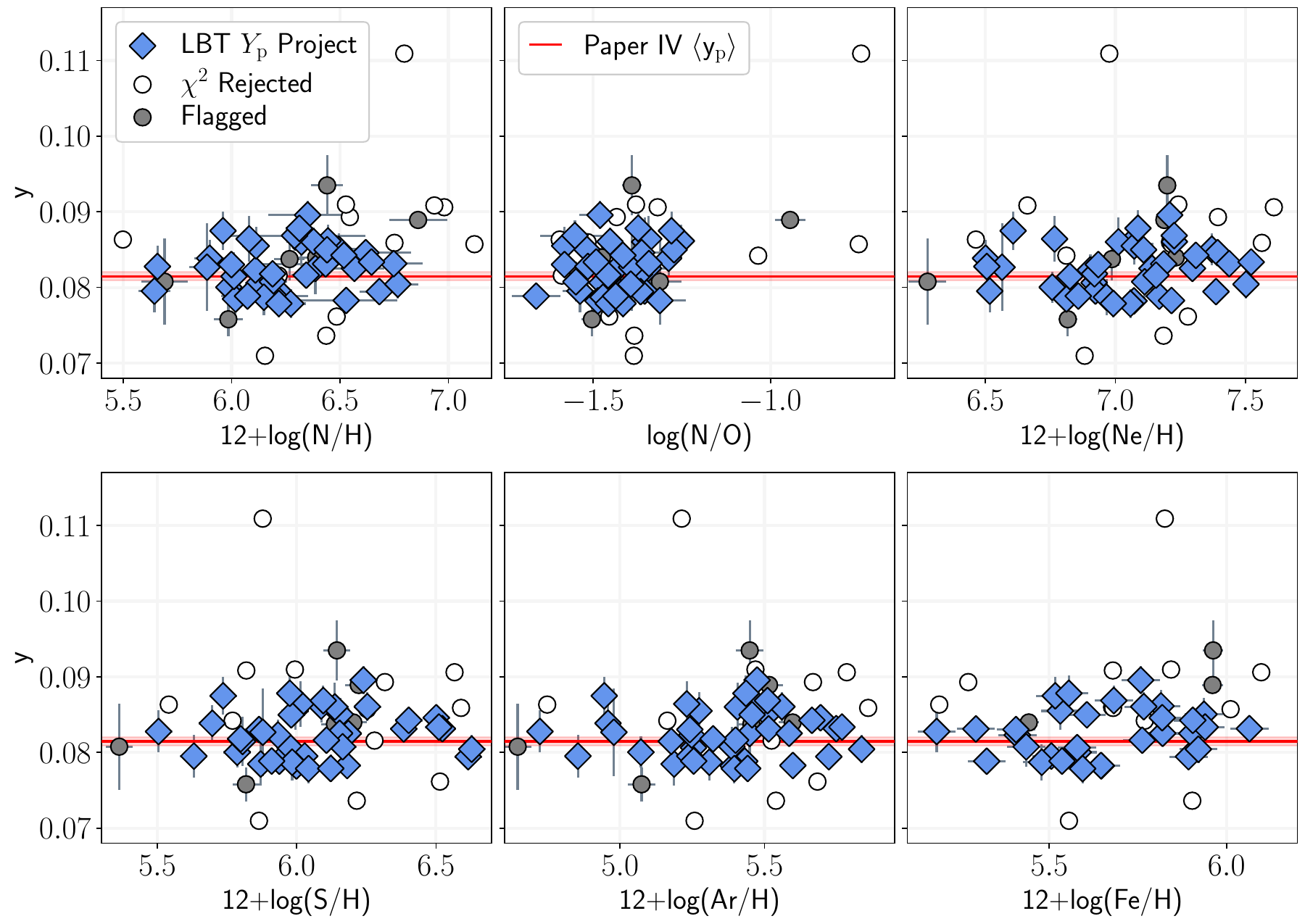}
   \caption{He/H number density, y, plotted against total and relative metal abundances. Nebulae that failed to meet the $\chi^2$ threshold or were flagged for anomalous model parameters derived in Paper IV are highlighted in white and gray circles, respectively. The red line is $\langle$y$_{\rm p}$$\rangle$ derived from the 15 qualifying targets with the lowest O/H. There is little evolution in y with any of the other metal abundances, but targets with anomalously high N/O are frequently rejected based on their $\chi^2$ or model parameters (top row, center panel). This suggests more complex physical processes, which are not accounted for in the He abundance model used to measure y, influence the derived N/O abundances.}
   \label{fig:yabun}
\end{figure*}

We compare y against a variety of other gas-phase abundances in Figure \ref{fig:yabun}, and we note that some targets with MODS spectra lack y measurements. As described in Papers I and IV, in some instances the MODS observations were used to assess the feasibility of detecting \ion{He}{1} $\lambda$10830 from LUCI observations, to verify if the target was metal poor, and to determine if the nebula had any physical properties that would disqualify it from He abundance analysis. For example, the galaxy WJ1205+4551 has no corresponding LUCI observations because the MODS spectra revealed broad components on some strong emission lines and intense \fev$\lambda$4227. These properties are related to gas physics or ionization conditions that are not captured in the model used in deriving the He/H abundances, so LUCI followup of WJ1205+4551 was de-prioritized due to the high likelihood that this object would be disqualified from the \yp\ measurement.

We include nebulae that were rejected based on the $\chi^2$ criteria from Paper IV as white circles, but no MCMC uncertainties on y were derived because the line fluxes were in poor agreement with the predictions from the He abundance model. Nebulae that were flagged and removed from the \yp\ sample based on modeled properties (high \ion{He}{1} optical depth, discrepant $T$(He) and \te\oiii, significant underlying stellar absorption, see Paper IV) are plotted as gray circles. The red line represents the weighted average value of the primordial He number density, $\langle$y$_{\rm p}$$\rangle$ $=$ 0.08146$\pm$0.00056, from the 15 most O-poor qualifying targets.

There is no significant correlation between y and the other elemental abundances. We find that the Pearson correlation coefficients for the six abundances plotted in Figure \ref{fig:yabun} all have $p$ $>$ 0.17, suggesting no statistically significant correlation. This supports the findings in Paper IV, where the slope of the linear regression to y vs.\ O/H derived from the same nebulae was statistically consistent with zero ($\Delta$y/$\Delta$(O/H) $=$ 8$\pm$16). The intrinsic dispersion in the He abundances ($\sigma_{{\rm y},int} = 0.00262 \pm 0.00048$, Paper IV) limits the ability to reliably measure a linear correlation with the other elements.

We investigate whether the nebulae with the lowest metal abundances reproduce the weighted average y$_{\rm p}$. Examining the sample of qualifying objects with He abundances, we recompute $\langle$y$_{\rm p}$$\rangle$ from the 15 objects with the lowest N/H, Ne/H, S/H, Ar/H, and Fe/H.
All $\langle$y$_{\rm p}$$\rangle$ agree with the value measured in Paper IV; the maximum and minimum $\langle$y$_{\rm p}$$\rangle$ is 0.08164 and 0.08143 from the nebulae with the lowest N/H and S/H, respectively. This agreement may be expected for N, Ne, S, and Ar since the abundance of these metals relative to O is roughly constant at low metallicity, so the lowest O/H correspond to the lowest N/H, Ne/H, etc. 
The Fe/H plotted in Figure \ref{fig:yabun} is measured in fewer targets and is not corrected for dust depletion, but the $\langle$y$_{\rm p}$$\rangle$ from the 15 most Fe-poor objects is 0.08147, in excellent agreement with $\langle$y$_{\rm p}$$\rangle$ from Paper IV. The recovery of $\langle$y$_{\rm p}$$\rangle$ supports the use of other elements besides O as metallicity tracers when measuring y$_{\rm p}$ \citep[e.g.,][]{fern2018}, and it is reassuring that the result for y$_{\rm p}$ is insensitive to the choice of metallicity tracer.

An intriguing trend is found when comparing y to the relative N/O abundance ratio. As shown in the top row, center panel of Figure \ref{fig:yabun}, three low-metallicity galaxies with enhanced N/O fail to meet the $\chi^2$ threshold to be considered for \yp. An additional galaxy (HS1028+3834) with elevated N/O meets the $\chi^2$ criteria but is flagged for a high \ion{He}{1} optical depth of $\tau$ $>$ 13. The remaining galaxy with elevated N/O relative to the primary plateau is WJ1205+4551, which, as discussed above, was de-prioritized for LUCI observations based on the characteristics of the MODS optical spectrum. In total, targets with elevated N/O relative to the low-metallicity plateau are poorly fit by the model designed to measure He$^+$.
One or more aspects of the physical scenario assumed by the model, that being photoionization by a simple stellar population, Case B recombination and \ion{He}{1} self-absorption, is not appropriate for the \ion{H}{1} and \ion{He}{1} line intensities and, simultaneously, manifests as very large N/O. Given this trend, the combination of O/H and N/O could be used to screen \citep[e.g.,][]{yanagisawa2026} for objects that have physics beyond the He abundance model assumed in Paper IV and, therefore, are not suitable for measuring y without significant systematic uncertainties. Expanding the range of model parameters \citep[e.g., allowing for higher electron densities and optical depths, see][]{berg2026} or including more physical processes may help better account for these systematics and provide reliable He abundance measurements in these peculiar galaxies.

\begin{figure*}[t]
   \centering
   \includegraphics[width=0.99\textwidth]{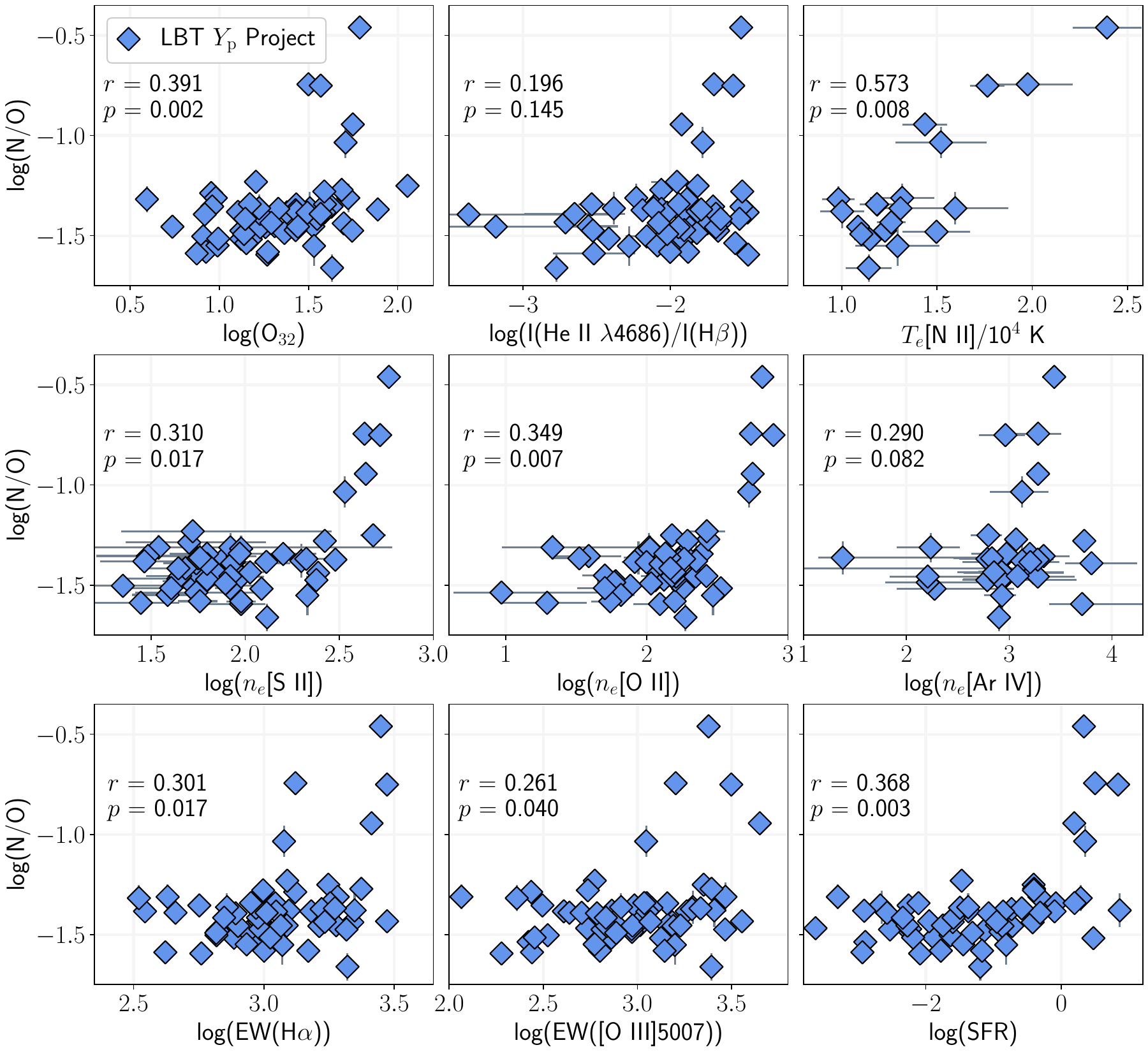}
   \caption{The relative abundance of N/O plotted against measured properties. From left to right: top row plots O$_{32}$, \ion{He}{2} $\lambda$4686/H$\beta$ intensity ratio and \te\nii, middle row plots \den\sii, \den\oii\ and \den\ariv, and bottom row plots EW(H$\alpha$), EW(\oiii$\lambda$5007), and SFR. The Spearman rank order correlation coefficients and $p$ values are provided in each panel. \den\sii, \den\oii, \te\nii, SFR, O$_{32}$, and EW(H$\alpha$) show the most significant correlations. However, none of the parameters offer a clear explanation of the mechanism responsible for the N-enhanced ISM.}
   \label{fig:no_all}
\end{figure*}

\section{N/O Enhanced Galaxies}\label{s:highno}

We now return to the five low metallicity galaxies with log(N/O) $>$ $-$1.1: J0519+0007, HS0837+4717, HS1028+3843, SHOC133, and WJ1205+4551. While they comprise a small fraction of the total sample ($\sim$10\% of nebulae with O/H $<$ 8.0 dex), local nebulae with this abundance pattern are relatively rare \citep[see][]{kobu1996,vanZ2006,berg2012,berg2019,izot2012,jame2015,bhattacharya2026b}. In contrast to surveys of local galaxies, observations of distant galaxies have revealed a population of bright, metal-poor, N/O enhanced galaxies \citep[e.g.,][]{marq2024,senchyna2024,topp2024,topping2025,berg2026b,ji2026}. These five galaxies, therefore, provide a rare opportunity to explore N/O enhancements in ISM with detailed constraints on the physical properties and chemical abundance patterns of other elements.
In this section, we examine the properties of these N/O enhanced galaxies, systematics related to their chemical abundance measurements, and briefly discuss potential causes for the elevated N/O.

\subsection{Galaxy Properties}

To illustrate the characteristics of these galaxies, Figure \ref{fig:no_all} plots log(N/O) against a variety of measured properties in the LBT \yp\ Project sample. Each panel includes the Spearman correlation coefficient and $p$ value. This is not an exhaustive list, but is meant to highlight potential mechanisms that could produce the N/O enhancement or bias measurements of N/O.

The N-enhanced galaxies are described by very high ionization conditions. All galaxies have O$_{32}$ $\geq$ 10 (top left panel) and nebular \ion{He}{2} $\lambda$4686 intensities $>$ 2\% the intensity of H$\beta$ (top middle panel). All but HS0837+4717 have significant detections of \fev$\lambda$4227. At such high ionization conditions, the simple assumption that N/O $\approx$ N$^+$/O$^+$ may break down due to the deficient O$^+$ in the ISM (see ICF predictions in Figure \ref{fig:n_icf}). However, this explanation is inconsistent with the trend of other highly ionized objects in the sample. While WJ1205+4551 has the largest O$^{2+}$/O ($\sim$0.979), the next highest ionization object (J1323$-$01325, O$^{2+}$/O $\sim$ 0.975, O$_{32}$ $=$ 39.2) has log(N/O) $=$ $-$1.251$\pm$0.023, which is only marginally above the primary N/O plateau. The significant correlation between log(N/O) and log(O$_{32}$) is likely an artifact of the dependence of both quantities on the \oii$\lambda$3727 line intensity. This is supported by the lack of significant correlation between log(N/O) and the \ion{He}{2} $\lambda$4686/H$\beta$ ratio, which traces the presence of very high-ionization gas. In short, while highly ionized gas is a characteristic of the N-enhanced galaxies, other nebulae with similar ionization conditions do not show significantly elevated N/O.

Another common characteristic of the N/O enhanced galaxies is their relatively high densities in the low-ionization gas. The middle row of Figure \ref{fig:no_all} reveals that the five galaxies have log(\den\sii) $>$ 2.5 and log(\den\oii) $>$ 2.7, among the highest in the sample. Recent studies of abundance patterns in local and high-\emph{z} galaxies have found a relation between N/O and \den\sii\ \citep{arel2025,martinez2025}, and the trends in Figure \ref{fig:no_all} are in qualitative agreement with these findings. The \oii$\lambda$3727 doublet is sensitive to collisional de-excitation, and if \den\ were underestimated then the O$^+$ derived from the \oii\ intensity would be biased low and result in a high N/O. However, we adopt the average of \den\sii\ and \den\oii\ as a fiducial density when calculating ionic abundances if \den\sii\ $>$ 10$^2$ cm$^{-3}$ \citep[see Paper II and][]{mend2023}. Consistent with the findings of \citet{arel2025}, no significant correlation is observed between N/O and \den\ariv; although the \ariv\ doublet is sensitive to higher densities, it may not appropriately reflect the density of the \oii\ emitting gas.

All five N-enhanced galaxies have significant detections of \nii$\lambda$5755 and a direct measurement of \te\nii, which ranges from 14400$\pm$1200 K in HS1028+3843 to 23900$\pm$1800 K in WJ1205+4551. There are fewer nebulae with direct \te\nii\ on account of low overall N abundance and high ionization conditions (see Paper II), but the \te\oiii\ and \te\nii\ in four of the N-enhanced galaxies disagree with the predictions from existing \te\ scaling relations (see Figure 8 in Paper II).

Interestingly, the 20 objects with direct \te\nii\ show a correlation between N/O and \te\nii\ (top right panel). The presence of this strong correlation is unexpected given the $j_{NII,6584}$/$j_{OII,3727}$ emissivity ratio's relatively weak sensitivity to \te. Fixing \den\ $=$ 10$^2$ cm$^{-3}$ and varying \te\ from 1.25-2.5$\times$10$^4$ K results in a roughly factor of 2 change in this emissivity ratio (similar results are obtained at \den\ $=$ 10$^3$ cm$^{-3}$). This translates to a 0.27 dex increase in log(N/O), notably less than N/O enhancement observed in the five galaxies relative to the N/O plateau. Instead, the presence of the auroral \nii$\lambda$5755 and a measured \te\nii\ that is significantly larger than \te\oiii\ \citep[at odds with \te\ scaling relations such as][]{camp1986,garn1992,roge2021} may indicate the presence of a complex physical structure or non-stellar ionizing sources within these low-metallicity galaxies.

The last row in Figure \ref{fig:no_all} plots the EW(H$\alpha$), EW(\oiii$\lambda$5007), and SFRs for the LBT \yp\ Project sample. For the latter, we first input an object's redshift, R.A.\ and Decl.\ into Cosmicflows-4 Distance-Velocity calculator \citep{kour2020,vala2024} to retrieve the distance to the object \citep[exceptions are the distances for Leo P and AGC198691, which are taken from][]{mcquinn2020}. The H$\alpha$ luminosity is calculated using this distance and the reddening-corrected H$\alpha$ intensity, which is then input into the \citet{kennicutt1998} relation to determine SFR. The N/O enhanced galaxies have very large EW(H$\alpha$) $>$ 1000 \AA\ and SFRs $>$ 1 M$_\odot$ yr$^{-1}$. This pattern is consistent with the N/O abundances in the high-SFR CLASSY galaxies \citep{arel2025} and high-\emph{z} JADES galaxies \citep{cameron2026}. To achieve such high EW(H$\alpha$) at low metallicity requires very young stellar populations \citep[$\lesssim$ 3 Myr, see][]{levesque2013,eldr2017,hawcroft2025}, although we note that systematics such as dust scattering, SFH, and aperture effects can complicate the inference of stellar population age from EW(H$\alpha$). Nevertheless, the young ages are consistent with the high SFRs on $\sim$10 Myr timescales implied from the H$\alpha$ luminosities.

\subsection{Systematics in N/O}

While we have discussed systematics such as the ICF and its impact on the N/O measurement, the physical conditions in the ISM could be biased and lead to an anomalously high N/O. 
Using HS0837+4717 as a test case, we now consider what change in  physical conditions is required to decrease the reported log(N/O) $=$ $-$0.74 to produce agreement with the primary plateau. Holding the \nii/\oii\ line intensity ratio and \den\ constant and varying \te\nii, log(N/O) $=$ $-$1.43 is only recovered in this galaxy at \te\nii\ $\approx$ 6900 K. This temperature is more than 10$^{4}$ K lower than inferred from the \nii$\lambda$5755/\nii$\lambda$6584 ratio, and such a low \te\nii\ is wholly inconsistent with the \te\oiii-\te\nii\ scaling relations at \te\oiii\ $=$ 19500$\pm$300 K. However, the calculation of \te\ depends on \den, where not accounting for significant collisional de-excitation of the \oiii$\lambda$5007 or \nii$\lambda$6584 would bias \te\oiii\ and \te\nii\ high \citep[e.g.,][]{martinez2025,berg2026b}.

\begin{figure}[t]
   \centering
   \includegraphics[width=0.47\textwidth]{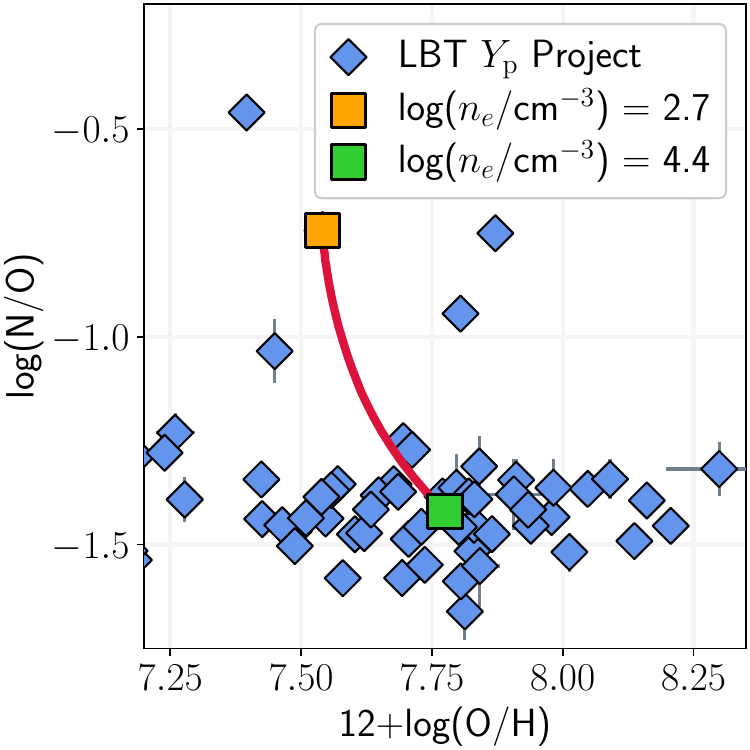}
   \caption{N/O vs.\ O/H, but the abundances in HS0837+4717 are highlighted in the orange square. The \te\ and abundances in this galaxy are calculated at a characteristic log($n_e$) $=$ 2.7 based on the \oii\ and \sii\ strong lines. The solid red line shows the effect of increasing the characteristic log(\den), resulting in increased O/H and decreased N/O. The abundance patterns implied by log(\den) $=$ 4.4 (green square) agree with the primary plateau observed in the LBT \yp\ Project. One explanation for the N/O enhanced galaxies is an underestimated $n_e$ in the ISM, although the \oii\ and \sii\ CEL ratios are inconsistent with this factor of $\sim$50 increase in density.}
   \label{fig:no_den}
\end{figure}

To illustrate this potential bias, we experiment with changing the characteristic density used to calculate \te\oiii, \te\nii, O$^+$, O$^{2+}$, and N$^+$ in HS0837+4717. Figure \ref{fig:no_den} plots the N/O-O/H abundances in the LBT \yp\ Project, where HS0837+4717's abundance patterns are calculated at log(\den/cm$^{-3}$) $=$ 2.7 and represented as the orange square. Increasing the characteristic \den\ and using the relevant line intensities to re-calculate \te\ and ionic abundances at the higher density, the red line plots the resulting changes to the abundance patterns in this galaxy. The green square represents the N/O and O/H abundances when the characteristic density is log(\den/cm$^{-3}$) $=$ 4.4, significantly greater than the critical densities of the \oii$\lambda\lambda$3726,29 emission lines. The total O/H increases by $\sim$0.25 dex, a significant change although the O/H remains $\sim$10\% of the solar abundance. In contrast, the relative N/O abundance decreases 0.7 dex and is now consistent with the primary plateau.

Clearly, not accounting for the presence of very high density gas when computing \te\ and ionic abundances has the potential to significantly bias the inferred chemical abundance patterns. The N/O-O/H abundance pattern in HS0837+4717 could be the result of an under-estimated \den, but this requires the density to be 50$\times$ larger than measured from the \sii\ and \oii\ lines. The same conclusion is reached when using the auroral-to-nebular \sii\ line ratio as a density tracer, and log(\den/cm$^{-3}$) $=$ 4.4 is more than an order of magnitude larger than the direct measurement of \den\ariv\ (see Figure \ref{fig:no_all}). Another potential bias is a poorly-constrained reddening, where an underestimated $E(B-V)$ would produce a larger \nii/\oii\ intensity ratio and N/O relative abundance. We examine the line intensities of high-order Balmer lines relative to H$\beta$ to assess whether the attenuation by dust has been under-predicted for the blue emission lines. In HS0837+4717, we measure I(H12)/I(H$\beta$) $=$ 0.032$\pm$0.002 and I(H11)/I(H$\beta$) $=$ 0.040$\pm$0.002, both in excellent agreement with the theoretical line ratios at the measured \te\nii\ and \den\sii\ (0.031 and 0.040, respectively).

Taken together, a large change to the physical conditions or dust attenuation in HS0837+4717 is required to produce an N/O abundance that agrees with the primary plateau\footnote{Similar conclusions are reached when repeating this exercise for the other N/O enhanced galaxies.}. The change in \te\ required at the adopted density of log(\den/cm$^{-3}$) $=$ 2.7 is not supported by the \te\ scaling relations, and the reddening-corrected \ion{H}{1} recombination line intensities agree with the theoretical ratios.
While a significantly larger characteristic density would change \te\ and ionic abundances sufficiently to move HS0837+4717 onto the primary plateau, this is also inconsistent with the direct density measurements. While optical CELs can underestimate \den\ \citep[e.g.,][]{mend2023,martinez2025}, it is unclear if the nearly two order of magnitude difference in $n_e$ would go undetected across all ionization zones. Lastly, this exercise assumes that N/O must be consistent with the primary plateau; if the true gas-phase N/O is marginally enhanced, then the required increase in \den\ or decrease in low-ionization zone \te\ is not as extreme.

\begin{figure*}[t]
   \centering
   \includegraphics[width=0.95\textwidth]{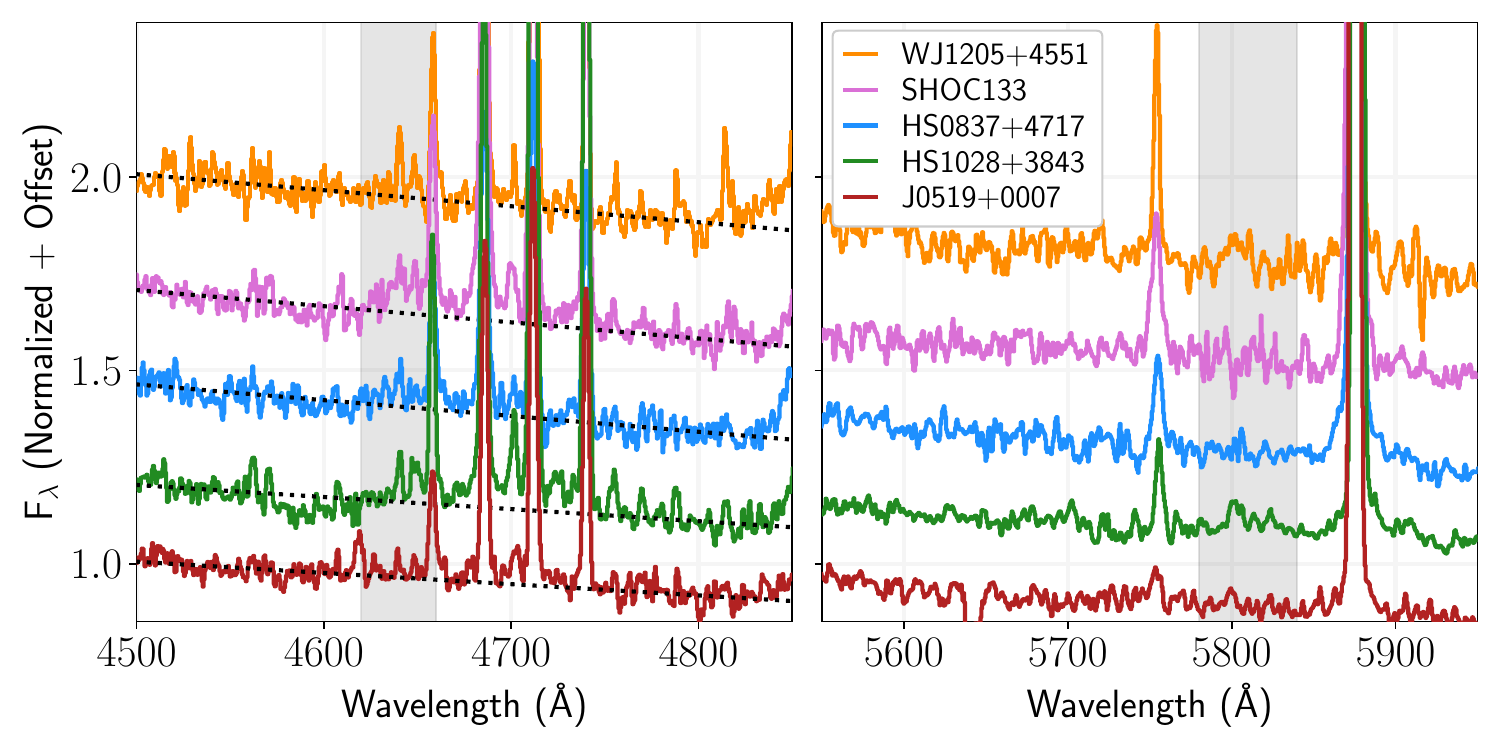}
   \caption{Portions of the LBT/MODS spectra of the five N/O enhanced galaxies, where the positions of the WR bumps in the blue (left panel) and red optical (right panel) are highlighted as gray shaded areas. Each spectrum is normalized and offset by an arbitrary amount to highlight the potential WR features. A linear continuum is plotted as a black dotted line in the left panel. While some broad \heii$\lambda$4686 is detected, most galaxies lack any emission near the blue bump, and the red WR features are absent. If WR stars are responsible for the N/O enhancement, then the rest-optical wind features are weak in the metal-poor galaxies.}
   \label{fig:no_wr}
\end{figure*}

\subsection{Potential Explanations for N/O Enhancement}

The N/O enhanced galaxies in the LBT \yp\ Project are experiencing significant star formation, have very high ionization conditions, relatively high densities in the low-ionization gas, and the surprising detection of auroral \nii\ at low metallicity. Relative to O, the other metal abundances in these galaxies are consistent with the rest of the sample: the difference between Ne/O, S/O, Ar/O, and Cl/O measured in the N/O enhanced galaxies and the averages reported in \S\ref{s:abun} are all less than 0.02 dex. However, four galaxies with LUCI NIR observations were rejected for He abundance analysis based on their poor agreement with the He abundance model or peculiar physical conditions.
We now consider scenarios that could produce large N abundance in the otherwise metal-poor ISM and/or bias the low-ionization emission.

Enhanced N/O could be the result of pollution from the N-rich winds of massive stars during the Wolf-Rayet (WR) phase. N/O enhancements have been observed in local \citep{walsh1987,kobulnicky1997,james2009,lopez-sanchez2010b} and high-\emph{z} galaxies \citep{mori2025,welc2025,berg2026b} with strong spectral features from WR winds. The duration of the WR phase is relatively short \citep[$<$ 1 Myr,][]{eldr2017}, although the fraction of massive stars in the WR phase is sensitive to the SFH. The relatively intense star formation and short ages implied for the N/O enhanced galaxies may be consistent with a short-lived WR phase.

In Figure \ref{fig:no_wr}, we plot the 1D spectra of each N/O enhanced galaxy around the location of the common broad emission features observed in WR nebulae. These features are often described as a ``blue bump", a combination of \ion{N}{3} $\lambda$4640, \ion{C}{3} $\lambda$4650, and \heii$\lambda$4686, and a ``red bump" from blended \ion{C}{3} $\lambda$5968 and \ion{C}{4} $\lambda$5808 \citep{lope2010}. The blue bump is produced by the WN class of WR stars, while the red bump is a feature of the WC class.
From the right panel, there is no clear excess emission at the position of the red bump (gray shaded area). This is to be expected since WN stars should be more prevalent than WC stars at low metallicities \citep{meynet2005,crowther2007,eldr2017,neugent2019}.
The only galaxy with a tentative excess around the red bump, HS1028+3843, shows a slightly broad component to \heii$\lambda$4686. In addition, HS0837+4717 and SHOC133 have a broad component to \heii$\lambda$4686, but there is only minor excess emission around 4640 that could be associated with \ion{N}{3}. Furthermore, J0519+0007 and WJ1205+4551 do not show broad \heii\ emission or blue WR features. The narrow, nebular \heii\ emission is detected in each galaxy and ranges from 1.2\% to 3.0\% the intensity of H$\beta$ (see Figure \ref{fig:no_all}). 

Although WR stars could produce elevated N/O in the ISM, the low-metallicity of the five galaxies results in ambiguous detections of the common optical WR features. 
An alternative explanation is enrichment from very massive stars (VMS, M/M$_\odot$ $>$ 100), which have been detected in the Tarantula Nebula in the Large Magellanic Cloud \citep[][]{crowther2016}. Due to strong convection and mixing, synthesized N can be brought to the surface of a VMS and ejected through strong stellar winds \citep{vink2023}. VMS have short lifetimes ($\sim$2 Myr) and generate a significant flux of H- and He$^+$-ionizing photons that can lead to intense nebular \heii\ emission \citep[e.g.,][]{hawcroft2025}. In these aspects, the impact of VMS and WR stars on the emission line fluxes, EWs, and chemical abundances are similar. \citet{martins2023,martins2025} proposed that the two scenarios can be distinguished with optical spectroscopy: VMS have broad and intense \heii, but often lack broad emission from \ion{N}{3}$\lambda$4640 and \ion{C}{4}$\lambda$5801 that are present in WN and WC stars.
The young implied ages for the N/O enhanced galaxies, their very high ionization conditions, minimal emission around 4640 \AA\ and 5801 \AA, and partially broad \heii\ are qualitatively consistent with ionization and enrichment from metal-poor WR stars or VMS.

Alternatively, the low ionization emission could be biased due to photoionization by non-stellar sources including radiative shocks or active galactic nuclei (AGN). These galaxies are consistent with predominant ionization from stars in common diagnostic diagrams such as those proposed by \citet{bald1981} (see Paper II), but a small contribution from harder ionizing sources would affect the relative emission line ratios of the low- and high-ionization zone ions. Particularly, these sources produce a significant flux of x-ray photons that can ionize He$^{+}$. Additionally, the path length of the high-energy photons is longer than low-energy photons and can produce large partially-ionized zones that enhance \nii, \oi, and \sii\ emission \citep{ho2008}.
If a portion of the \nii\ and \oii\ emission is not associated with the \hii\ region and, instead, is produced in high-temperature and/or density gas ionized by radiative shocks or AGN, then the N/O abundances derived under the assumption of only photoionization could be significantly biased.

The presence of other ionizing sources in the N/O enhanced galaxies may explain the inability of the He abundance model to recover the observed H and He line intensities. Disentangling the relative contribution of ionization from stars (including WR and VMS), shocks, or an AGN may be feasible with photoionization models that consider combinations of these sources \citep[e.g.,][]{izot2021,marconi2024,chisholm2024}. 
Partial ionization from an accreting intermediate mass black hole (IMBH) is an intriguing explanation. Nebulae ionized by an IMBH and stars can mimic low-metallicity galaxies in the \citet{bald1981} line ratio diagrams \citep[e.g.,][]{richardson2022,chisholm2024}, which may explain the optical line ratios of the five N/O enhanced galaxies.
Additional observations are required to confirm these different ionization or chemical enrichment scenarios. Rest-UV spectroscopy of the stellar continuum and features such as broad \heii$\lambda$1640 and \ion{N}{4}$\lambda$1486 P Cygni profiles can help differentiate between the WR and VMS scenarios in conjunction with optical spectroscopy \citep{martins2023}. Such spectra would also provide coverage of UV CELs from \ion{N}{3} and \ion{N}{4}, which could better constrain the ISM density structure and N abundance.
UV and MIR spectra contain other tracers of complex ionizing sources, thereby revealing whether the low-ionization emission lines are unsuitable for N/O abundances. Particularly, features such as [\ion{Ne}{5}]$\lambda$14.3$\mu$m and the dust continuum will be useful in assessing the presence of an AGN in these intriguing galaxies.

\section{Conclusions}\label{s:conclusions}

In this paper, we discuss the metal abundance patterns in local, low-metallicity nebulae in the LBT \yp\ Project. The high-quality spectroscopy allows for a measurement of up to 13 ionic abundances from 8 elements, in addition to the He abundances derived in Paper IV. We use the homogeneous sample to discuss the nucleosynthetic production of these elements and the expected abundance trends in the metal-poor ISM at \emph{z} $=$ 0. To summarize our key findings:
\begin{enumerate}
    \item With the \te, \den, and the metal CELs from Paper II, we measure ionic abundances in three different ionization zones. We compare available ICFs for the commonly-measured elements N, Ne, S, and Ar (Figures \ref{fig:n_icf} and \ref{fig:alpha_icf}), adopting the common assumption that N$^+$/O$^+$ $=$ N/O, the ICF(Ne$^{2+}$) of \citet{dors2013}, and the \citet{izot2006} ICFs for S and Ar.
    \item Owing to the selection criteria of the LBT \yp\ Project sample and the depth of the MODS spectra, we are able to constrain the total abundances of other trace elements in the ISM. Emission lines from Fe$^{2+}$ and Fe$^{4+}$ are detected, and we use the ICF of \citet{rodr2005} to measure Fe/H in 47 nebulae. Cl is measured in 41 targets, where 26 have significant detections of \cliv$\lambda$7540,8046. Assuming negligible Cl$^+$ in the high-ionization ISM, we show that ICFs for Cl systematically under-predict the amount of Cl$^{3+}$ in the ISM (Figure \ref{fig:cl_icf}). In addition, we calibrate an empirical Cl ICF for use when Cl$^{3+}$ is the only ionic abundance measured in the ISM (Equation \ref{eq:icfcl3}). We report the first detections of \kiv$\lambda$6102 in extragalactic \hii\ regions and star-forming galaxies, and use the available ICF(K) to derive total K abundances.
    \item The N/O abundances in the low-metallicity nebulae are constant, in agreement with prior findings in other local metal-poor \hii\ regions. The primary plateau measured from the LBT \yp\ Project sample is $-$1.43$\pm$0.10 dex (Figure \ref{fig:lno}), although there are five peculiar galaxies with N/O significantly enhanced above the primary plateau. Combining the sample with the metal-rich CHAOS \hii\ regions, we calibrate a new expression relating the gas-phase N/O and O/H (Figure \ref{fig:lno_full}, Equation \ref{eq:no2}) which reproduces the solar N/O abundance at solar O/H.
    \item The $\alpha$ element ratios Ne/O, S/O, and Ar/O all show no evolution with metallicity and agree with the solar ratios (Figure \ref{fig:pure_alphao}). This abundance pattern is related to a common nucleosynthetic origin from massive stars and a fixed yield, but S and Ar have a non-negligible production from Type Ia SNe. An evolution in S/O and Ar/O has been observed in high-\emph{z} galaxies that are experiencing rising and abbreviated SFHs. The abundance ratios in the LBT \yp\ Project nebulae, therefore, are related to the extended SFHs experienced by local low-mass, metal-poor galaxies.
    \item Ratios of the odd-Z elements Cl and K to O show no evolution with metallicity and are marginally sub-solar (Figure \ref{fig:adj_alpha}). These elements behave similarly to $\alpha$ process elements, where their production is regulated by the abundance of $\alpha$ elements in massive stars. The ten K/O abundances are the first of their kind in terms of their low bulk metallicity (12+log(O/H) $<$ 8.0), and the average K/O agrees with measurements in more metal-rich PNe. The gas-phase Cl and K abundances in the metal-poor nebulae are significantly larger than predicted by chemical evolution models, necessitating a change to the nuclear reaction rates that control the production of odd-Z elements.
    \item Fe/O in the LBT \yp\ Project nebulae shows a strong anti-correlation with O/H, as has been observed in other extragalactic nebulae (Figure \ref{fig:feo}). This trend may be related to dust depletion of Fe that becomes more significant with increasing metallicity. Using an empirical correction that considers the O and N abundances, the average Fe/O is approximately $-$0.8 dex below the solar ratio. If this low gas-phase Fe abundance reflects the stellar Fe content, then the resulting harder radiation spectra of the low-metallicity stars contribute to the high-ionization conditions in the LBT \yp\ Project sample.
    \item We compare the He mass fraction to total abundances of different metals and find that $\langle$y$_{\rm p}$$\rangle$ is consistent with that derived from the weighted average of the 15 most O-poor nebulae from Paper IV (Figure \ref{fig:yabun}). Interestingly, galaxies with elevated N/O are preferentially rejected or flagged for removal from the He abundance analysis. This pattern argues that there are physical conditions in these galaxies not considered in the physical model for deriving He abundances.
    \item We discuss the properties of the low-metallicity, N/O enhanced galaxies, finding that these galaxies have high low-ionization zone densities, ionization conditions, EWs (consistent with stellar populations with age $<$ 3 Myr), and SFRs compared to the full sample (Figure \ref{fig:no_all}). These properties are also observed in high-\emph{z} star-forming galaxies with elevated N/O. We examine systematics that can bias N/O measurements such as an underestimated \den. We conclude with an overview of potential sources that could enhance N/O, including low-metallicity WR stars or VMS. Alternatively, ionization mechanisms like shocks or accretion onto an IMBH that are unaccounted for could contribute to the low-ionization emission lines and change the derived N/O. The detection of high-ionization lines such as \heii$\lambda$4686, in addition to the departure from the He abundance model predictions, argue for a more complex ISM (multiple ionizing sources, densities, etc.) that may impact the N/O measurements.
\end{enumerate}

In addition to a precision measurement of the primordial He abundance, the LBT \yp\ Project has delivered a comprehensive set of chemical abundance patterns in more than 60 low-metallicity nebulae in the local Universe. Such a homogeneous sample will prove useful for interpreting the chemistry of similarly low-metallicity galaxies observed at high-\emph{z} with JWST. Additionally, the average abundance patterns measured here can be used as starting points for grids of photoionization models tailored for the metal-poor ISM.

New observations and future work stand to meaningfully increase our understanding of the physics and chemistry of the low-metallicity ISM. It is clear that new Cl and K ICFs are required for high-ionization galaxies, which will improve the precision of the gas-phase Cl/O and K/O reported in \S\ref{s:oddz}. Prior studies have used Cl to interpret the enrichment of odd-Z elements in the gas, but this work demonstrates the ability to measure the abundance of K in high-ionization extragalactic nebulae. Additional rest-optical observations of bright star-forming galaxies and extragalactic \hii\ regions will provide a statistical sample of K abundances capable of addressing the buildup odd-Z elements over a wider range of metallicity, which will be useful for constraining the input reaction rates for chemical evolution models.

Novel studies are possible with the combination of the LBT/MODS and LUCI spectra with rest-UV and IR spectroscopy. Coverage into the FUV provides access to stellar wind features that may be associated with WR or VMS, thereby assessing the presence of these stars in galaxies with extreme ionization conditions or enhanced N/O. UV CELs of N can be detected and resolved with HST/COS, allowing for an independent measurement of N/O and the complex density structure in the ISM \citep[e.g.,][]{ming2022,martinez2025}. Similarly, JWST/MIRI spectroscopy covers fine-structure lines of Ne, S, Cl, Ar, and K, including those from ionization states not observable in the optical. These fine-structure lines allow for the derivation of new empirical ICFs to validate predictions from photoionization model grids, and can resolve the detailed \te\ and \den\ structure in the different ionization zones. These studies and more will be possible with the MIRI spectroscopy of the LBT \yp\ Project nebulae acquired as part of the Mid-Infrared Abundances of Gaseous Environments (MIRAGE, PID:7973, PI: Rogers) survey.

\begin{acknowledgements}
The authors would like to thank Stephanie Monty and Chiaki Kobayashi for their helpful discussion concerning odd-Z element production and enrichment. This work was supported by funds provided by NSF Collaborative Research Grants AST-2205817 to RWP, AST-2205864 to EDS, and AST-2205958 to EA.
Our team workshop at OSU in July 2024 was sponsored in part by OSU's Center for Cosmology and AstroParticle Physics (CCAPP).

This work is based on observations made with the Large Binocular Telescope. The LBT is an international collaboration among institutions in the United States, Italy and Germany. LBT Corporation Members are: The University of Arizona on behalf of the Arizona Board of Regents; Istituto Nazionale di Astrofisica, Italy; LBT Beteiligungsgesellschaft, Germany, representing the Max-Planck Society, The Leibniz Institute for Astrophysics Potsdam, and Heidelberg University; The Ohio State University, and The Research Corporation, on behalf of The University of Notre Dame, University of Minnesota and University of Virginia. Observations have benefited from the use of ALTA Center (alta.arcetri.inaf.it) forecasts performed with the Astro-Meso-Nh model. Initialization data of the ALTA automatic forecast system come from the General Circulation Model (HRES) of the European Centre for Medium Range Weather Forecasts.

This research used the facilities of the Italian Center for Astronomical Archive (IA2) operated by INAF at the Astronomical Observatory of Trieste.

EDS, EA, DAB, NSJR, and JHM would like to acknowledge and thank Stanley Hubbard for his generous gift to
the University of Minnesota that allowed the University to become a member of the LBT collaboration.

\end{acknowledgements}

\facilities{LBT (MODS), LBT (LUCI)}
\software{
\texttt{astropy} \citep{astr2013, astr2018, astr2022},
\texttt{jupyter} \citep{kluy2016},
\texttt{modsIDL} \citep{crox2019},
\texttt{modsCCDRed} \citep{pogg2019},
\texttt{PypeIt} \citep{proc2020},
\texttt{PyNeb} \citep{luri2012,luri2015L},
\texttt{numpy} \citep{harr2020},
\texttt{scipy} \citep{virt2020}}

\startlongtable
\begin{deluxetable*}{lcccccccc}  
\tablecaption{LBT \yp\ Project Chemical Abundances \label{t:abun}}
\tablewidth{\columnwidth}
\tabletypesize{\footnotesize}
\tablehead{
   \colhead{Property} & 
   \colhead{AGC198691} & 
   \colhead{DDO68} & 
   \colhead{HS0029+1748} &
   \colhead{HS0122+0743} &
   \colhead{HS0134+3415} &
   \colhead{HS0811+4913} &
   \colhead{HS0837+4717}}
\startdata
\hline
\hline
O$^{+}$/H$^+$ ($\times$10$^5$)   & 0.24$\pm$0.05 & 0.27$\pm$0.03 & 1.43$\pm$0.28 & 0.53$\pm$0.07 & 0.51$\pm$0.25 & 1.54$\pm$0.33 & 0.21$\pm$0.07 \\
O$^{2+}$/H$^+$ ($\times$10$^5$)   & 0.77$\pm$0.11 & 1.25$\pm$0.07 & 9.74$\pm$0.32 & 3.27$\pm$0.11 & 6.42$\pm$0.19 & 7.98$\pm$0.25 & 3.26$\pm$0.12 \\
12+log(O/H)   & 7.002$\pm$0.050 & 7.181$\pm$0.022 & 8.048$\pm$0.016 & 7.580$\pm$0.015 & 7.841$\pm$0.020 & 7.979$\pm$0.019 & 7.541$\pm$0.017 \\
\vspace{-0.15cm} \\
N$^{+}$/H$^+$ ($\times$10$^6$)   & 0.12$\pm$0.02 & 0.08$\pm$0.01 & 0.61$\pm$0.07 & 0.14$\pm$0.01 & 0.25$\pm$0.07 & 0.57$\pm$0.08 & 0.38$\pm$0.08 \\
12+log(N/H)   & 5.690$\pm$0.107 & 5.644$\pm$0.075 & 6.682$\pm$0.091 & 5.999$\pm$0.060 & 6.528$\pm$0.235 & 6.545$\pm$0.098 & 6.797$\pm$0.156 \\
log(N/O)   & $-$1.312$\pm$0.061 & $-$1.537$\pm$0.049 & $-$1.366$\pm$0.033 & $-$1.581$\pm$0.023 & $-$1.312$\pm$0.073 & $-$1.434$\pm$0.036 & $-$0.744$\pm$0.046 \\
\vspace{-0.15cm} \\
Ne$^{2+}$/H$^+$ ($\times$10$^6$)   & 1.59$\pm$0.26 & 2.85$\pm$0.19 & 21.72$\pm$0.85 & 7.61$\pm$0.30 & 15.06$\pm$0.53 & 21.69$\pm$0.82 & 8.73$\pm$0.37 \\
Ne ICF   & 1.19$\pm$0.03 & 1.15$\pm$0.01 & 1.12$\pm$0.01 & 1.13$\pm$0.01 & 1.09$\pm$0.02 & 1.14$\pm$0.02 & 1.08$\pm$0.01 \\
12+log(Ne/H)   & 6.277$\pm$0.072 & 6.517$\pm$0.029 & 7.387$\pm$0.018 & 6.934$\pm$0.018 & 7.216$\pm$0.017 & 7.394$\pm$0.018 & 6.976$\pm$0.019 \\
log(Ne/O)   & $-$0.725$\pm$0.088 & $-$0.664$\pm$0.037 & $-$0.661$\pm$0.024 & $-$0.646$\pm$0.023 & $-$0.625$\pm$0.026 & $-$0.585$\pm$0.026 & $-$0.564$\pm$0.025 \\
\vspace{-0.15cm} \\
S$^{+}$/H$^+$ ($\times$10$^7$)   & 0.45$\pm$0.05 & 0.50$\pm$0.03 & 2.62$\pm$0.21 & 1.06$\pm$0.06 & 1.34$\pm$0.26 & 2.12$\pm$0.19 & 0.50$\pm$0.07 \\
S$^{2+}$/H$^+$ ($\times$10$^7$)   & 1.44$\pm$0.23 & 2.45$\pm$0.38 & 21.51$\pm$0.77 & 4.44$\pm$0.15 & 5.87$\pm$0.29 & 12.04$\pm$0.55 & 3.16$\pm$0.27 \\
S ICF   & 1.22$\pm$0.12 & 1.45$\pm$0.10 & 1.71$\pm$0.17 & 1.33$\pm$0.05 & 2.11$\pm$0.60 & 1.46$\pm$0.13 & 2.06$\pm$0.37 \\
12+log(S/H)   & 5.363$\pm$0.068 & 5.631$\pm$0.064 & 6.615$\pm$0.046 & 5.865$\pm$0.021 & 6.181$\pm$0.127 & 6.317$\pm$0.042 & 5.878$\pm$0.085 \\
log(S/O)   & $-$1.639$\pm$0.084 & $-$1.550$\pm$0.068 & $-$1.432$\pm$0.048 & $-$1.715$\pm$0.026 & $-$1.659$\pm$0.128 & $-$1.662$\pm$0.046 & $-$1.663$\pm$0.086 \\
\vspace{-0.15cm} \\
Cl$^{2+}$/H$^+$ ($\times$10$^8$)   & \nodata & \nodata & 2.17$\pm$0.48 & \nodata & 0.77$\pm$0.06 & 1.40$\pm$0.11 & 0.50$\pm$0.11 \\
Cl$^{3+}$/H$^+$ ($\times$10$^8$)   & \nodata & \nodata & 0.46$\pm$0.10 & 0.41$\pm$0.07 & 0.90$\pm$0.05 & 0.81$\pm$0.05 & \nodata \\
Cl ICF   & \nodata & \nodata & 1.00 & 2.41$_{-0.10}^{+0.10}$ & 1.00 & 1.00 & 1.31$_{-0.19}^{+0.42}$ \\
12+log(Cl/H)   & \nodata & \nodata & 4.420$_{-0.081}^{+0.081}$ & 3.994$_{-0.074}^{+0.074}$ & 4.223$_{-0.021}^{+0.021}$ & 4.345$_{-0.023}^{+0.023}$ & 3.842$_{-0.118}^{+0.153}$ \\
log(Cl/O)   & \nodata & \nodata & $-$3.628$_{-0.083}^{+0.083}$ & $-$3.586$_{-0.076}^{+0.076}$ & $-$3.618$_{-0.029}^{+0.029}$ & $-$3.634$_{-0.030}^{+0.030}$ & $-$3.699$_{-0.119}^{+0.154}$ \\
\vspace{-0.15cm} \\
Ar$^{2+}$/H$^+$ ($\times$10$^7$)   & 0.40$\pm$0.13 & 0.61$\pm$0.10 & 4.14$\pm$0.14 & 1.13$\pm$0.05 & 1.92$\pm$0.08 & 3.41$\pm$0.14 & 0.92$\pm$0.07 \\
Ar$^{3+}$/H$^+$ ($\times$10$^7$)   & \nodata & \nodata & 1.16$\pm$0.10 & 0.61$\pm$0.03 & 2.03$\pm$0.07 & 1.25$\pm$0.06 & 0.70$\pm$0.05 \\
Ar ICF   & 1.12$\pm$0.03 & 1.18$\pm$0.03 & 0.99$\pm$0.00 & 1.00$\pm$0.00 & 1.00$\pm$0.01 & 1.00$\pm$0.00 & 1.01$\pm$0.00 \\
12+log(Ar/H)   & 4.646$\pm$0.144 & 4.854$\pm$0.073 & 5.722$\pm$0.014 & 5.241$\pm$0.015 & 5.598$\pm$0.012 & 5.668$\pm$0.014 & 5.213$\pm$0.023 \\
log(Ar/O)   & $-$2.357$\pm$0.153 & $-$2.327$\pm$0.076 & $-$2.326$\pm$0.022 & $-$2.338$\pm$0.021 & $-$2.243$\pm$0.023 & $-$2.311$\pm$0.024 & $-$2.328$\pm$0.029 \\
\vspace{-0.15cm} \\
K$^{3+}$/H$^+$ ($\times$10$^8$)   & \nodata & \nodata & \nodata & \nodata & 0.33$\pm$0.04 & 0.21$\pm$0.05 & \nodata \\
K ICF$^{a}$   & \nodata & \nodata & \nodata & \nodata & 2.17$\pm$1.26 & 2.19$\pm$1.26 & \nodata \\
log(K/Ar)   & \nodata & \nodata & \nodata & \nodata & $-$1.456$\pm$0.110 & $-$1.431$\pm$0.140 & \nodata \\
12+log(K/H)   & \nodata & \nodata & \nodata & \nodata & 4.142$\pm$0.111 & 4.236$\pm$0.141 & \nodata \\
log(K/O)   & \nodata & \nodata & \nodata & \nodata & $-$3.698$\pm$0.113 & $-$3.743$\pm$0.142 & \nodata \\
\vspace{-0.15cm} \\
Fe$^{2+}$/H$^+$ ($\times$10$^6$)   & \nodata & \nodata & 0.13$\pm$0.03 & 0.05$\pm$0.01 & 0.04$\pm$0.02 & 0.04$\pm$0.01 & 0.06$\pm$0.01 \\
Fe$^{4+}$/H$^+$ ($\times$10$^6$)   & \nodata & \nodata & \nodata & \nodata & \nodata & \nodata & \nodata \\
Fe ICF$^{b}$   & \nodata & \nodata & 0.77$\pm$0.01 & 0.78$\pm$0.01 & 0.73$\pm$0.03 & 0.79$\pm$0.01 & 0.72$\pm$0.02 \\
12+log(Fe/H)$^{c}$   & \nodata & \nodata & 5.893$\pm$0.114 & 5.406$\pm$0.101 & 5.646$\pm$0.261 & 5.272$\pm$0.143 & 5.826$\pm$0.171 \\
log(Fe/O)$^{c}$   & \nodata & \nodata & $-$2.155$\pm$0.121 & $-$2.174$\pm$0.104 & $-$2.195$\pm$0.272 & $-$2.707$\pm$0.152 & $-$1.715$\pm$0.177 \\
\enddata
\tablecomments{Chemical abundances measured from the LBT/MODS spectra. Rows represent direct ionic abundances, ICFs, and total and relative abundances for the elements O (Rows 1-3), N (4-6), Ne (7-10), S (11-15), Cl (16-20), Ar (21-25), K (26-30), and Fe (31-35). Table truncated at seven targets, full version available in machine readable format. $^a$K ICF is used to measure K/Ar, which is then converted to K/H and K/O. $^b$Fe ICF is used to measured Fe/O, which is converted to Fe/H. $^c$Total and relative Fe abundances are not corrected for dust depletion.}
\end{deluxetable*}



\bibliographystyle{aasjournalv7}
\bibliography{yp_total_bib}

\clearpage

\end{document}